\documentclass{aa}  

\usepackage{graphicx}
\usepackage{txfonts}
\usepackage{xcolor}

\newcommand{\srg}{\textit{SRG}}
\newcommand{\xmm}{\textit{XMM-Newton}}

\begin{document}

   \title{Tempestuous life beyond $R_{500}$: \\X-ray view on the Coma cluster with SRG/eROSITA.}

   \subtitle{I. X-ray morphology, recent merger, and radio halo connection.}
   
   \author{E. Churazov \inst{1,2} 
   \and
   I. Khabibullin \inst{2,1}
   \and
   N.Lyskova \inst{2,1}
   \and
   R.Sunyaev \inst{2,1}
   \and
   A.M.Bykov \inst{3}
   }
\institute{Max Planck Institute for Astrophysics, Karl-Schwarzschild-Str. 1, D-85741 Garching, Germany 
\and 
Space Research Institute (IKI), Profsoyuznaya 84/32, Moscow 117997, Russia
\and
Ioffe Institute, Politekhnicheskaya st. 26, Saint Petersburg 194021, Russia}

   \date{Received ; accepted }

  \abstract
{This is the first paper in a series of studies of the Coma cluster using the \textit{SRG}/eROSITA X-ray data obtained in course of the Calibration and Performance Verification observations. The data cover $\sim3^\circ\times 3^\circ$ area around the cluster with a typical exposure time of more than 20~ks. The stability of the instrumental background and operation  of the \textit{SRG} Observatory in the scanning mode provided us with an excellent data set for studies of the diffuse emission up to a distance of $\sim 1.5R_{200}$ from the Coma center. In this study, we discuss the rich morphology revealed by the X-ray observations (also in combination with the SZ data) and argue that the most salient features can be naturally explained by a recent (on-going) merger with the NGC~4839 group. In particular, we identify a faint X-ray bridge connecting the group with the cluster, which is convincing proof that NGC~4839 has already crossed the main cluster. The gas in the Coma core went through two shocks, first through the shock driven by NGC~4839 during its first passage through the cluster some Gyr ago, and, more recently, through the "mini-accretion shock" associated with the gas settling back to quasi-hydrostatic equilibrium in the core. After passing through the primary shock, the gas should spend much of the time in a rarefaction region, where radiative losses of electrons are small, until the gas is compressed again by the  mini-accretion shock. Unlike "runway" merger shocks, the mini-accretion shock does not feature a rarefaction region downstream and, therefore, the radio emission can survive longer. Such a two-stage process might explain the formation of the radio halo in the Coma cluster.
}


   \maketitle
%


\section{Introduction}
~~~~Coma cluster (Abell 1656) is one of the best-studied clusters in all energy bands. A combination of its mass ($M_{500}\sim 6\,10^{14}\,M_\odot$; \citealt{2013A&A...554A.140P}) and proximity ($z=0.0231$) makes it an attractive target -- bright and well resolved -- for many case studies. For instance, the presence of Dark Matter in the Coma cluster was suggested by Fritz Zwicky in 1933 based on velocity measurements of member galaxies \citep{1933AcHPh...6..110Z}. The Coma cluster was one of the very first clusters detected in X-rays \citep[e.g.][]{1966PhRvL..17..447B, 1972ApJ...178..309F}. In radio band, it became the first cluster, where a radio halo \citep[e.g.][]{1957ApJ...126..585S, 1959Natur.183.1663L, 1970MNRAS.151....1W}  and a radio relic \citep[e.g.][]{1979ApJ...233..453J, 1981A&A...100..323B, 1982Ap.....18..107G} have been detected. In microwaves, it became the first nearby cluster detected \citep{1995ApJ...449L...5H} via Sunyaev-Zeldovich (SZ) effect \citep{1972CoASP...4..173S}. Evidence for diffuse emission in the extreme ultraviolet band from the Coma cluster has been reported \citep{1996Sci...274.1335L}. X-ray observations of the Coma cluster provided robust evidence \citep{1993Natur.366..429W} that the baryon fraction in the cluster is lower than that derived from the Big Bang nucleosynthesis in a flat cosmology without Dark Energy. Recently, analysis of statistical properties of the X-ray surface brightness fluctuations allowed \cite{2019NatAs...3..832Z} to measure effective viscosity in the turbulent magnetized gas of the intracluster medium in Coma outskirts.

While initially classified as a relaxed cluster  \citep[e.g.][]{1982AJ.....87..945K}, it later became 
recognized as a merger \citep[e.g. ][]{1987ApJ...317..653F, 1992A&A...259L..31B, Neumann.et.al.2003}.  Recent studies of the Coma cluster in X-rays revealed a number of substructures, which are plausibly connected to the recent merger(s), 
namely, a prominent X-ray excess to the West of the Coma center \citep{Neumann.et.al.2003, 2013ApJ...775....4S}, an excess to the East, elongated in the East-West direction \citep{1997ApJ...474L...7V, Neumann.et.al.2003, 2013ApJ...775....4S}, the NGC 4839 group merging with Coma \citep[e.g.][]{Burns.et.al.1994,Neumann.et.al.2001, Lyskova2019, 2020A&A...634A..30M, 2020MNRAS.497.3204M}, and high density `arms' in the Coma core \citep{2012MNRAS.421.1123C, 2013Sci...341.1365S, 2019NatAs...3..832Z, 2020A&A...633A..42S}.
Similarly, remarkable progress with characterizing the Coma cluster has been achieved in the optical and radio bands \citep[][among others]{Adami.et.al.2005, 2020A&A...634A..30M, 2011MNRAS.412....2B, 2020arXiv201108856B}.

Deep X-ray observations covering the entire cluster, including the outskirts, with the good spectral and angular resolution, are indispensable to address the following problems:
\begin{itemize}
\item Validate and refine the scenario for a merger between the Coma cluster and the NGC~4839 group.
\item Provide accurate characterization of the "baryonic boundary" of the cluster, well beyond $R_{500}$, where the gas joins the cluster for the first time.
\item Examine a connection between X-ray emission and radio sources Coma C  and 1253+275, which themselves are the archetypes of clusters' radio halos and radio relics, respectively.
\end{itemize}
Here (paper I) we focus on the merger scenario and its connection to the radio halo using the data from the eROSITA telescope onboard the \srg~ observatory. 

The \srg~ X-ray observatory \citep{2021arXiv210413267S}  was launched on July 13, 2019,  from the Baikonur cosmodrome. It carries two wide-angle grazing-incidence X-ray telescopes, eROSITA \citep{2021A&A...647A...1P} and Mikhail Pavlinsky ART-XC telescope \citep{2021arXiv210312479P}, operating in the overlapping energy bands of 0.3–10 and 4–30 keV, respectively. On Dec. 13th, 2019, upon completion of the commissioning, calibration, and performance verification phases, \srg~ started its all-sky X-ray survey from a halo orbit around the Sun–Earth L2 point, which will consist of 8 repeated 6-month-long scans of the entire sky. Here we report the first (preliminary) results on the Coma cluster obtained in the course of the calibration and performance verification observations by \srg/eROSITA. The large field of view of eROSITA combined with operation in a special scanning mode provided an exceptionally uniform and deep X-ray image covering $\sim$10 sq.degrees centered on the Coma cluster.


Throughout the paper, we assume a flat $\Lambda$CDM model with $\Omega_m = 0.3$, $\Omega_{\Lambda}$=0.7, $H_0 = 70$ km/s/Mpc. At the redshift of the Coma cluster of $z = 0.0231$, one arcmin corresponds to 27.98 kpc. We assume $r_{500}=(47\pm 1)$ arcmin \citep{2013A&A...554A.140P} and $r_{200} \simeq 1.5 r_{500} \simeq 70$ arcmin for the concentration parameter $c\simeq4$ typical for massive galaxy clusters \citep[e.g.][among others]{2016MNRAS.457.4340K}. For all radial profiles, we fix the Coma center  at (RA, DEC) = (12$^h$59$^m$47$^s$, +27$^{\circ}$55'53"), as it is done in \cite{2013A&A...554A.140P}.

\begin{figure*}
\centering
\includegraphics[angle=0,bb= 50 170 550 665,width=1.7\columnwidth]{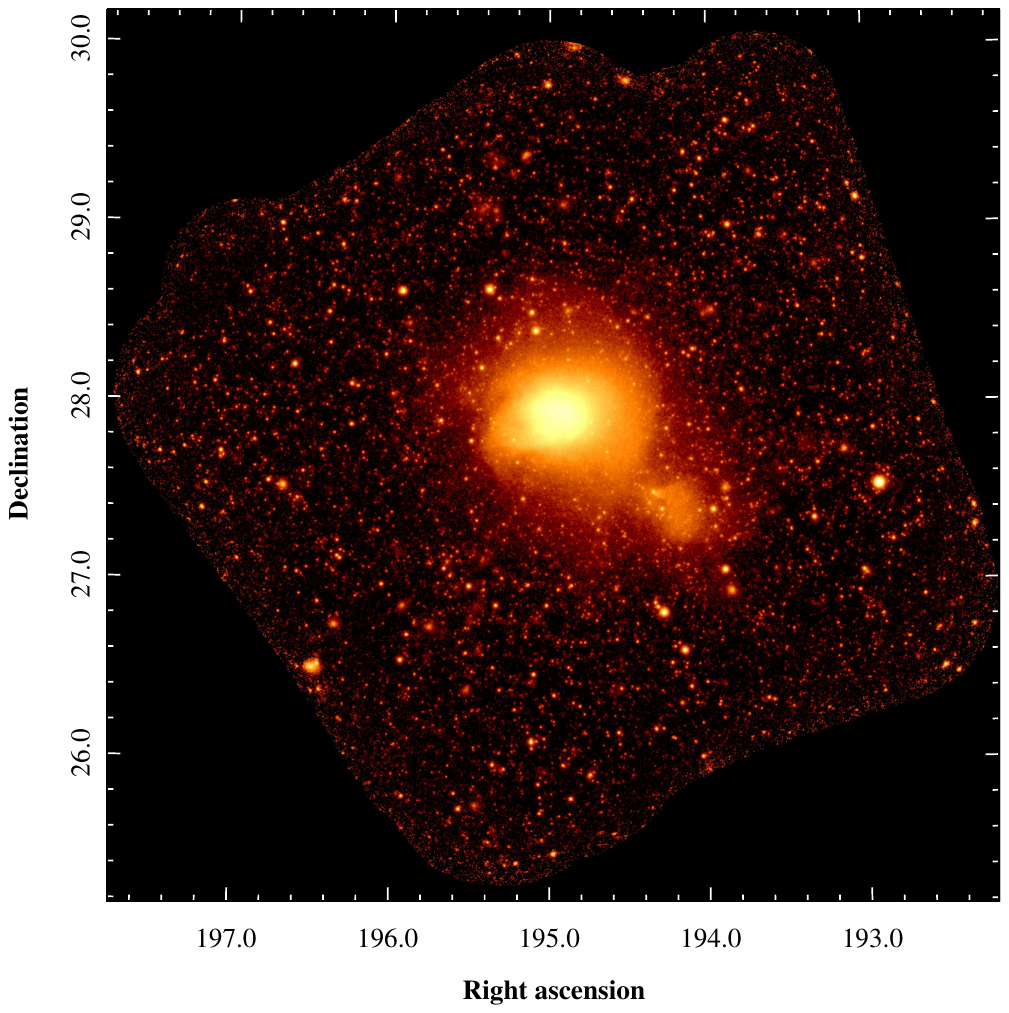}
\caption{X-ray image (background subtracted, exposure corrected) of the Coma cluster field in the 0.4-2 keV band accumulated during two scanning observations. The image is $\sim$6 degrees on a side, corresponding to 10~Mpc at the distance of the cluster, with the logarithmic color-coding on a scale spanning 5 orders of magnitude.}
\label{fig:coma_ximage}
\end{figure*}

\section{Observations and data handling}

The \srg\, observations of the Coma cluster were performed in two parts, on December 4-6, 2019 and June 16-17, 2020. The observatory was in a scanning mode (not to be mixed with the survey mode) when a rectangular region of the sky is scanned multiple times to ensure uniform exposure of the region. In both observations, the telescope axis was scanning across the $\sim 3\times 3^\circ$ field, albeit with a slight shift in the position angle between the two sessions. Here we report on \srg/eROSITA data obtained during these observations.

In order to produce the calibrated event lists, the raw data were reduced using the \texttt{eSASS} software. Based on the analysis of the light-curves, the data were cleaned from obvious artifacts. After the data cleaning, the accumulated exposure is fairly uniform over the $\sim 3^\circ\times 3^\circ$ field, reaching $\sim 23\,{\rm ks}$ per point at the center.


The detector intrinsic background was estimated using the data from the all-sky survey collected during the periods when the filter wheels of the telescope modules were in the "CLOSED" position, i.e. when the telescope detectors were not exposed to the X-ray emission of astrophysical origin. 
The total background (intrinsic detector background + astrophysical background) was estimated using observations of deep fields, taking into account the fraction of the resolved sources. Since the limiting flux of the resolved sources varies across the studied field (due to the contribution of the Coma cluster emission and variations in the exposure time at the edges of the field), the total background was interpolated (see Appendix~\ref{app:bg}) assuming Log\,N-Log\,S curve in the 0.5-2 keV band derived from \textit{Chandra} deep fields \citep{2017ApJS..228....2L}. 

For science analysis, only X-ray events within $28'$ from the center of the detector were used. Since the scanning mode is largely similar to the survey mode, we use the Point-Spread-Function (PSF) of the telescope estimated via stacking many compact sources in the all-sky survey data. It turns out that a simple $\beta$-model (see Appendix~\ref{app:psf}) provides a reasonable approximation for the shape of the PSF core. 

In this study, we use the data from all 7 eROSITA telescope modules for imaging, while the spectral analysis is limited to 5 telescopes equipped with the on-chip filter. These telescopes have a very similar response and, as a first approximation, the data can be directly co-added (as opposed to a more rigorous joint analysis of spectra obtained by individual telescopes). For the remaining 2 telescopes, we defer the analysis for future publications, when the quality of the spectral response calibration will allow the joint analysis to be performed. We reiterate here that some aspects of the eROSITA calibration are still preliminary, especially when dealing with hot clusters like Coma. We, therefore, adopted a conservative approach and report here only the results, which are robust against these uncertainties. Minor changes, especially related to the accurate temperature measurements are still possible and will be reported elsewhere. 

We note in passing that observations of the Coma cluster happened at the time of the Solar minimum. As a result, time variations of the detector particle background turned out to be small \citep[see][]{2021arXiv210312479P,2021arXiv210413267S} and the Solar Wind Charge Exchange (SWCX) emission is expected to be minimal, too. The impact of the latter component was found to be important for the analysis of the XMM-Newton data on the Coma cluster \citep[e.g.,][]{2008ApJ...680.1049T}. The location of the SRG at L2 point already eliminates the contribution of near-Earth regions to SWCX, leaving the heliosphere and the He-focusing cone as possible sites for the production of SWCX. The analysis of the SWCX as seen by the SRG observatory will be reported elsewhere, Here we only note that the data accumulated during two Coma observations separated by $\sim$6 months agree well, effectively providing good constraints on the (variable) SWCX flux.

\begin{figure}
\centering
\includegraphics[angle=0,trim= 0mm 0cm 0mm 0cm,width=1\columnwidth]{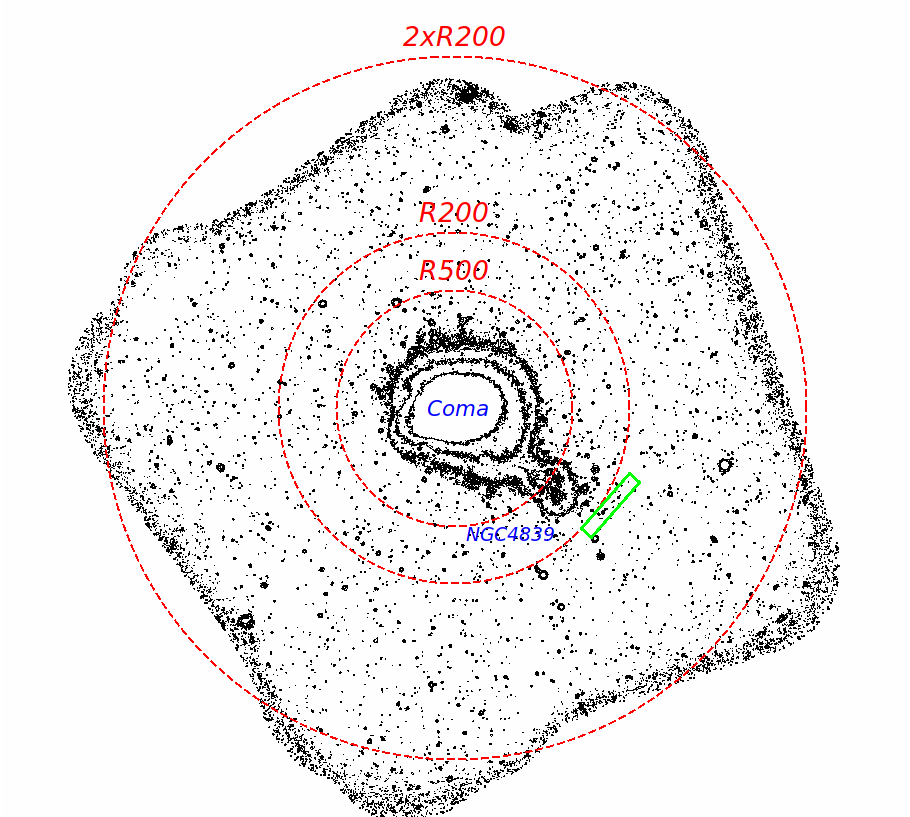}
\caption{Sketch of the Coma field with three circles showing estimated $R_{500}$, $R_{200}$, and $2\times R_{500}$. The NGC4839 group is approximately at $R_{500}$ ($\sim 1.3$~Mpc).
The green box marks the brightest part of the radio relic (radio source 1253+275), which is just outside $R_{200}$ ($\sim 2.1$~Mpc from the Coma center). }
\label{fig:coma_sketch1}
\end{figure}

\section{X-ray images}
The 0.4-2 keV band image of the full field, (detector) background subtracted, exposure corrected, and smoothed with a kernel corresponding to the telescope PSF is shown in Fig.~\ref{fig:coma_ximage} (with logarithmic color-coding). Such a smoothing kernel is chosen to optimize the sensitivity to faint point sources when the background dominates the signal.

A finding chart for the studied field is shown in Fig.~\ref{fig:coma_sketch1}, where black contours show the X-ray surface brightness. Three dashed circles indicate $R_{500}$, $R_{200}$ and $2\times R_{200}$ of the Coma cluster. It follows, that the data cover the region up to $\sim 1.5\times R_{200}$ extremely well. The extension of X-ray contours at $\sim R_{500}$ to the SW is due to the NGC~4839 group, which is in process of merging with the main cluster \citep[e.g.][ among many others]{1988A&A...199...67M, Neumann.et.al.2001}. This extension is aligned with the direction of a prominent filament of galaxies \citep[e.g.][]{2020A&A...634A..30M}. Farther out in the same direction, at $\sim R_{200}$, is located the famous radio relic source 1253+275.

A multitude (some 5000) of compact and diffuse sources is seen in the image, most of them being background active galactic nuclei (AGN). The sensitivity to point sources is a few $10^{-15}\;{\rm erg\,cm^{-2}\,s^{-1}}$ outside the very central region, where the diffuse emission from Coma dominates, and the very edges, where the exposure drops. The analysis of background AGNs and distant clusters, as well as foreground Galactic objects, in the Coma field will be reported in separate publications.

\begin{figure}
\centering
\includegraphics[angle=0,trim= 0mm 0cm 0mm 0cm,width=1\columnwidth]{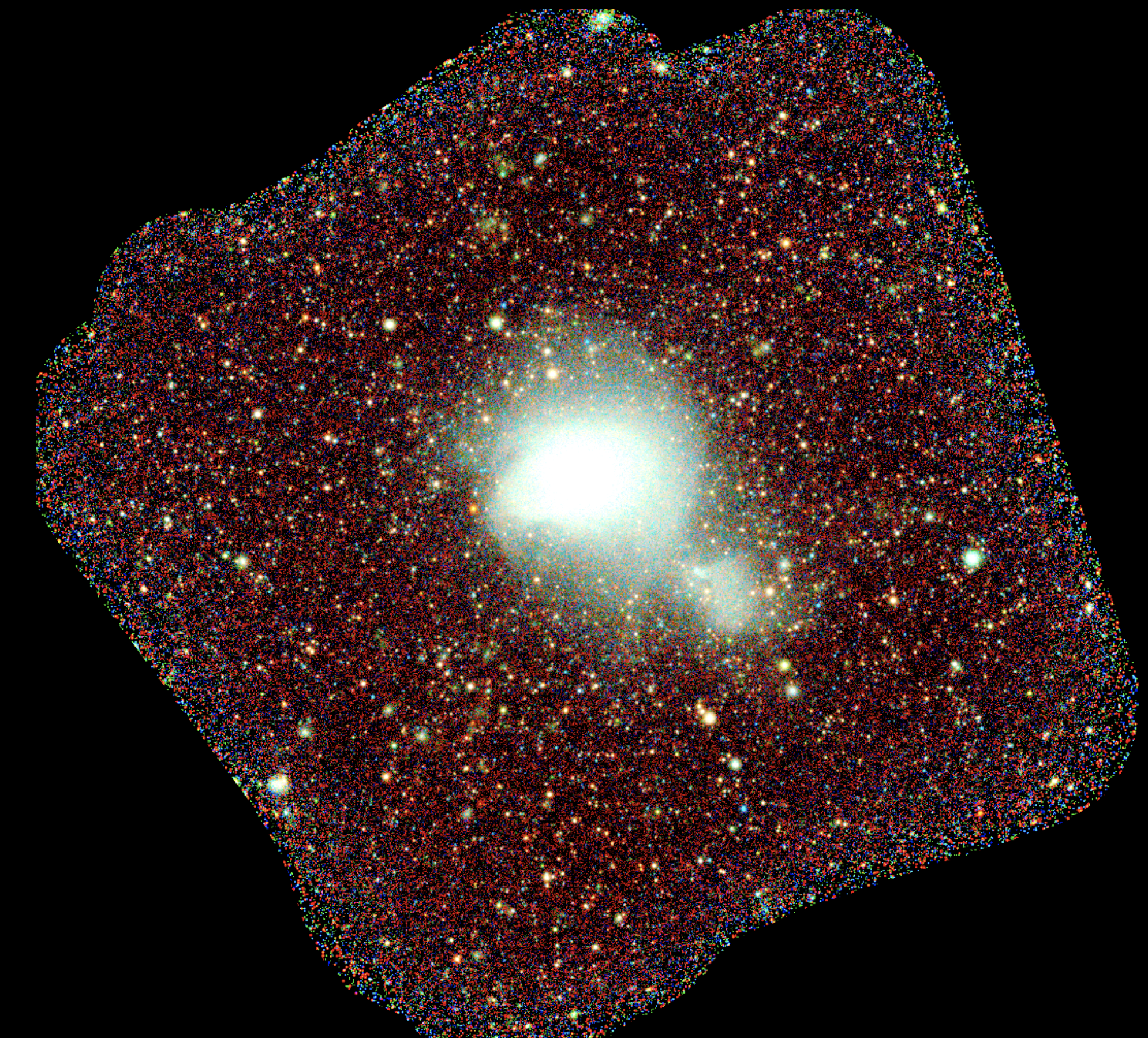}
\caption{Pseudo-color RGB image of the Coma cluster field based on a combination of 0.3-0.6, 0.6-1.0, and 1.0-2.3 keV images. The diffuse emission, which is dominated by the Milky Way contribution appears red in this image. Distant AGNs and a few galactic sources (including stars) span full range of colors. Clusters of galaxies, including the Coma cluster itself, appear as extended white objects in this image.}
\label{fig:coma_rgb}
\end{figure}

The pseudo-color RGB version of the X-ray image is shown in Fig.~\ref{fig:coma_rgb}. It is based on a combination of the 0.3-0.6, 0.6-1.0, and 1.0-2.3 keV images derived from the same data set. The colors were tuned to emphasize the emission from galaxy clusters. The overall reddish color of the image is due to the soft diffuse emission, in particular the OVII line, to which the Milky Way makes the dominant contribution. In this map, AGNs appear as bluish or yellowish objects, while galaxy clusters, including the Coma cluster itself, the NGC~4839 group, and more distant background clusters have a distinct white color.    

\begin{figure*}
\centering
\includegraphics[angle=0,bb= 50 170 550 665,width=1\columnwidth]{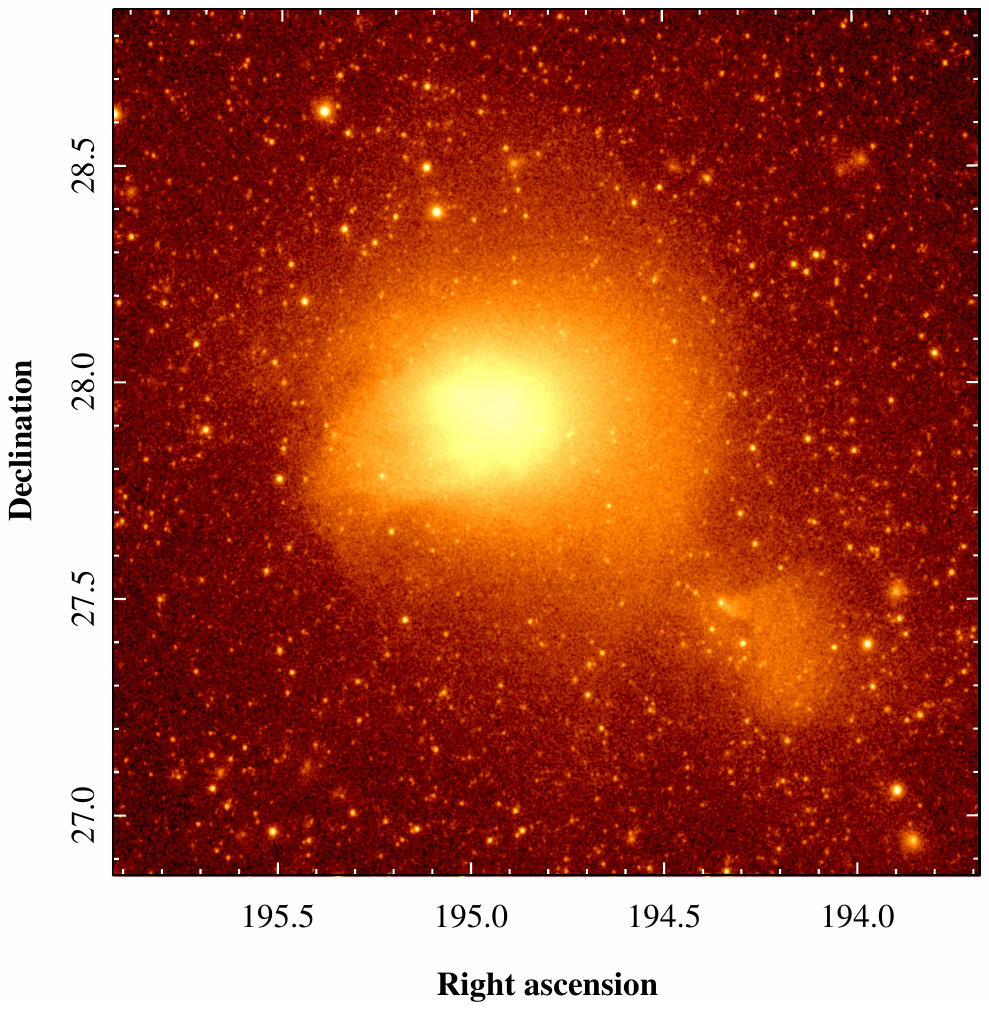}
\includegraphics[angle=0,bb= 50 170 550 665,width=1\columnwidth]{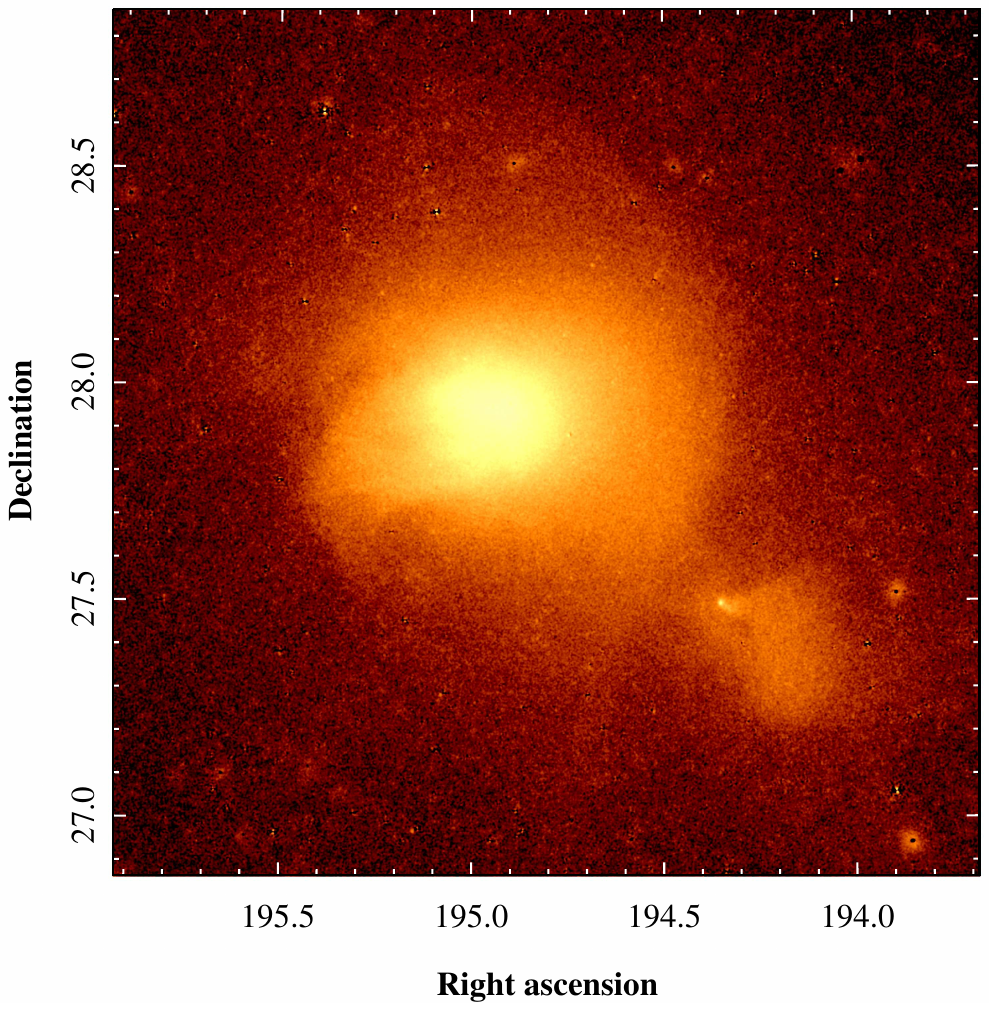}
\caption{X-ray image of the central $\sim2\times2$ degrees of the field showing 0.4-2 keV surface brightness on a a logarithmic scale. The left panel shows the original image without subtraction of the multitude of foreground, background and Coma-related sources. The right panel shows the map of the diffuse emission obtained after modeling and subtraction of the confidently detected point-like and mildly-extended sources using the $\beta$-profile approximation of the PSF described in Appendix \ref{app:psf}.  }
\label{fig:coma_central}
\end{figure*}

One can readily detect and subtract point (or mildly-extended) sources from the image in order to get a better view of the large scale distribution of Coma's diffuse emission. An image of the diffuse emission in the central $\sim2\times2$ degrees of the field, corresponding to $\sim 3.2$ Mpc, is shown in Fig.~\ref{fig:coma_central}.

All these images demonstrate rich morphology of the Coma cluster on Mpc scales, including a number of sharp "edges" close to $R_{500}$. The qualitative study of these structures is the primary focus of this paper. 

\subsection{Radial X-ray surface brightness profile}

\begin{figure}
\centering
\includegraphics[angle=0,trim= 0mm 5cm 0mm 3cm,width=1\columnwidth]{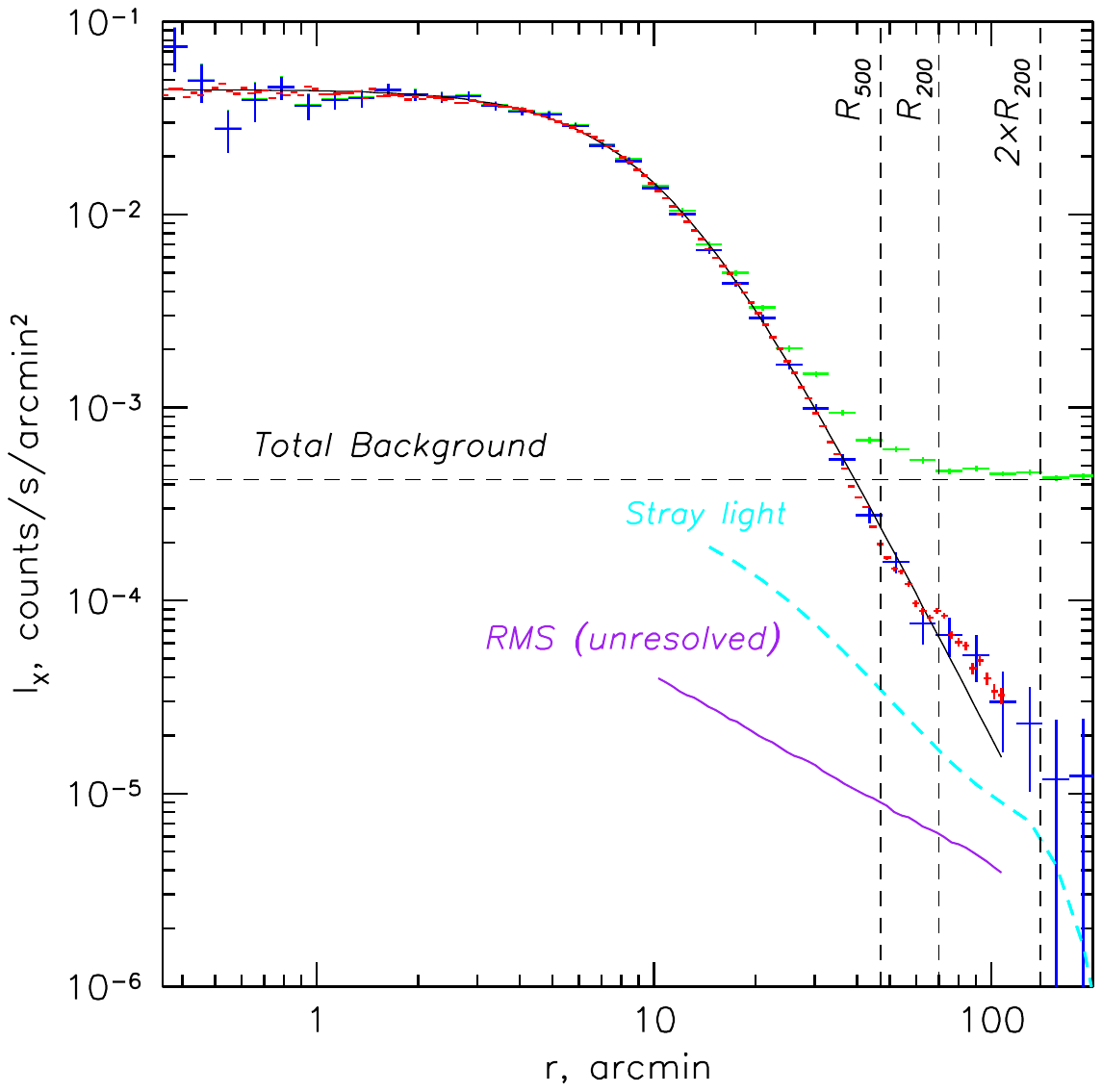}
\caption{Radial surface brightness profile of the Coma cluster in the 0.4-2 keV band. The surface brightness is normalized per single (out of 7) eROSITA telescopes. The $90^\circ$ wedge in the SW direction, which contains the NGC~4839 group, has been excluded from the analysis. The green points show the raw surface brightness (including all types of the background) as measured in the all-sky survey. The blue points are the same data after excising compact sources and removing different types of the residual background and foreground (both instrumental and astrophysical), and subtracting the estimated contribution of the stray light. The red points show similarly processed data from the data set discussed in this paper. For comparison, the thin black line shows a $\beta-$model with the co-radius $r_c=10'$ and $\beta=0.73$. Three vertical lines show  $R_{500}$, $R_{200}$, and $2\times R_{200}$. The dashed horizontal line shows the total background in the 0.4-2~keV band. The cyan curve shows the estimated level of the stray light, which was removed from the profiles. At the moment of writing, the exact shape and normalization of this component are still uncertain. Finally, the purple line shows the estimated level of flux variations associated with the Poisson fluctuations of the number of unresolved sources (in the outer regions where the Coma contribution is subdominant). 
 }
\label{fig:coma_radial}
\end{figure}

The radial profile of the X-ray surface brightness in the 0.4-2 keV band is shown in Fig.~\ref{fig:coma_radial}. A $90^\circ$ wedge to the SW, which contains the NGC~4839 group, has been excluded from this analysis.
The green points correspond to the total observed surface brightness profile. For comparison, the estimated total background surface brightness (detector intrinsic background plus astrophysical background) is shown with the dashed horizontal line. The red data points correspond to the surface brightness after excising sources and removing various components of the background, including the estimated contribution of the "stray-light", which is associated with photons scattered by the mirrors only once (as opposed to the nominal two-scatterings scheme in the Wolter I mirrors).

The radial profile of the stray-light component, which extends to some $3^\circ$ from the source, was estimated using observations of the brightest compact objects in the all-sky survey. The X-ray image (after subtraction of the detector intrinsic background) was been convolved with the stray-light profile, yielding the estimated contribution of the singly-scattered photons to the observed radial profile\footnote{Note that formally singly-scattered photons should be removed from the observed profile before the convolution. Here we ignore this second-order effect.}. This estimated contribution is shown in Fig.~\ref{fig:coma_radial} with the cyan dashed line. Since the X-ray signal from the bright sources used to derive the stray-light profile suffer from the pile-up effect, an accurate estimate of the stray-light contribution is problematic. We, therefore, defer the discussion of the X-ray radial profile at the largest radii to subsequent studies. In the present study, we focus on the data within $R_{200}$, where the importance of the stray-light component is less critical. 

Despite the remaining uncertainty in the stray-light magnitude, it is clear that with the present data set, the X-ray surface brightness profile can be traced all the way to $\sim 100'$, corresponding to $\sim 1.4\times R_{200}$. At this distance, the X-ray surface brightness is $5\,10^{4}$ times smaller than in the core of the cluster. For comparison, the purple line in Fig.~\ref{fig:coma_radial} shows the estimated root-mean-square (RMS) of the X-ray flux variations associated with the pure Poisson noise of the number of unresolved CXB sources using the known Log~N-Log~S taken from \cite{2017ApJS..228....2L}. Clearly, this is not the main source of uncertainty in the present data set.

In appendix \ref{app:rosat} we compare the radial profile extracted from the eROSITA all-sky survey data with the 0.4-2.4 keV profile obtained by \textit{ROSAT}. The good agreement of these profiles suggests that the background subtraction and the correction for the stray-light contribution do not introduce any major biases.

\subsection{Flat-fielded X-ray image}

\begin{figure*}
\centering
\includegraphics[angle=0,trim= 0mm 0cm 0mm 0cm,width=1.9\columnwidth]{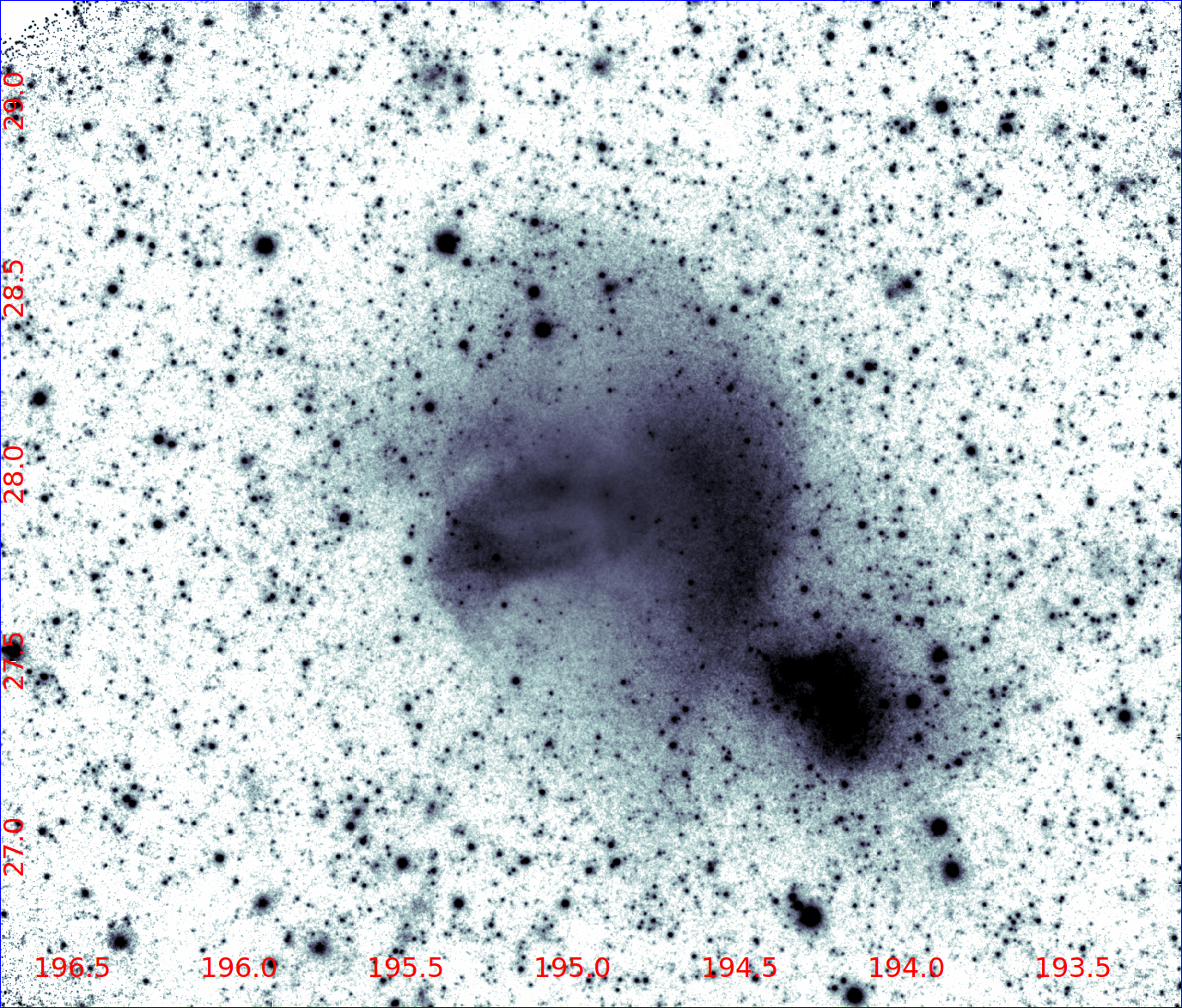} 
\caption{Flat-fielded $3.5\times 2.4$ degrees X-ray image of the Coma cluster. Flat-fielding procedure (see Eq.~\ref{eq:flat}) emphasizes departures from a smooth purely radial profile. The most spectacular is the bow-like sharp feature to the West from the core, which extends over $\sim 3$~Mpc from the North to the South. To the East from the core, there is another very sharp feature ($\sim$500~kpc in size), which turns out to be a contact discontinuity (see below). The NGC4839 (a dark spot to the SW from the core) appears to be connected with the main cluster via a rather faint "bridge". In addition, one can spot a number of fainter filaments/extensions, which will be discussed in the forthcoming studies.  
}
\label{fig:coma_flt}
\end{figure*}


As shown in the previous section, the traceable X-ray surface brightness varies by almost four orders of magnitude from the core to the outskirts. To cope with such a large dynamic range and to emphasize faint non-radial structures in the X-ray image, it is useful to artificially suppress the brightness of the core. A simple $\beta$-model with $r_c=10.4'$ and $\beta=0.73$ provides a reasonable first-order approximation to the observed profile. A  division by the $\beta$-model would flat-field the image, but it would also boost the noise in the cluster outskirts. We, therefore, divided the image by the following function 
\begin{eqnarray}
f(r)=1+\frac{c}{\left [1+\left (r/r_c \right )^2 \right ]^{3\beta-\frac{1}{2}}},
\label{eq:flat}
\end{eqnarray}
where $c$ is set to 30. With this definition, the division of the image by $f(r)$ at $r\ll r_c$ is equivalent to the division by the beta model. At large $r$, the value of $f(r)$ approaches unity, i.e. the image is unchanged. The parameter $c$ controls the transition radius and, implicitly, the level of the core brightness suppression relative to the outskirts, which is a factor $c+1=31$ in our case. The resulting image is shown in Fig.~\ref{fig:coma_flt}. 


This image reveals some of the features, which appear less prominently in the original image.  Some of these features have already been noticed in the ROSAT, XMM-Newton and Chandra images \citep[e.g.][]{1997ApJ...474L...7V, 1999ApJ...513..690D, 1999ApJ...527...80W, Neumann.et.al.2003, 2013Sci...341.1365S,2013ApJ...775....4S,2020MNRAS.497.3204M}, although in the flattened eROSITA image these features can now be traced over their full extent \footnote{We note in passing that the choice of the flat-fielding procedure is ambiguous and another procedure would lead to a slightly different appearance of the image.}. The most spectacular is a bow-like sharp feature to the West from the core, which is believed to be a shock (hereafter "W-shock"). In the eROSITA image, the W-shock boundary can be traced over $\sim$3~Mpc. Another sharp feature is seen to the East from the core, which turns out to be a contact discontinuity (see below).  

In addition, there are several new features that can be seen in this image. In particular, these include
\begin{itemize}
\item a faint "bridge" connecting the main cluster and the NGC4839 group;
\item an extended diffuse region to the west from the main cluster (better seen in the smoothed image, see \S\ref{sec:xsz});
\item several faint filaments extending to the South barely visible in the image;
\item a trace of an extended (in the North-South direction) surface brightness edge to the East from the core, which is reminiscent of a Bullet cluster shock \citep[e.g.][]{2002ApJ...567L..27M,2019A&A...628A.100D} ahead of a contact discontinuity.  
\end{itemize}
We discuss the origin of some of these features in the subsequent sections.


\subsection{Edges in X-ray image}
\label{sec:xsubstructure}
The original and flattened X-ray images shown above  feature a number of long but sharp "edges". Some of these structures are Mpc-long, while in the transverse direction the surface brightness changes over some 10 kpc or less. There are several plausible explanations for the X-ray appearance of such structures \citep[see, e.g.][for a review]{2007PhR...443....1M}, namely 
\begin{itemize}
\item a shock front, when the pressure is changing across the edge 
\item a contact discontinuity, when the pressure is continuous across the edge
\end{itemize}
In the X-ray data, the shocks reveal themselves as structures that have brighter (denser) and hotter gas on the same side of the interface, while the opposite is true for the contact discontinuities.   

In regard to contact discontinuities, one can further distinguish several cases
\begin{itemize}
\item discontinuity separating the Coma gas from the gas brought into the main cluster by the merger (could be still bound to the infalling halo or already stripped from it)
\item a displaced lump of the main cluster gas, which was brought (by motions) into contact with another lump with different entropy (also from the main cluster)
\item a contact discontinuity arising due to shock crossing \citep[e.g.][]{2010MNRAS.408..199B,2020MNRAS.494.4539Z}
\item a contact discontinuity associated with the variations of the fractional contribution of cosmic rays and magnetic fields to the total pressure.
\end{itemize}
The latter type of contact discontinuities is more relevant for cool core clusters, where bubbles of relativistic plasma created by a central AGN displaces the gas, while the first three could be present in a merging cluster, e.g. in Coma.

\begin{figure*}
\centerline{
\includegraphics[trim= 0cm -5cm 0cm 0cm, clip=t, width=0.49\linewidth]{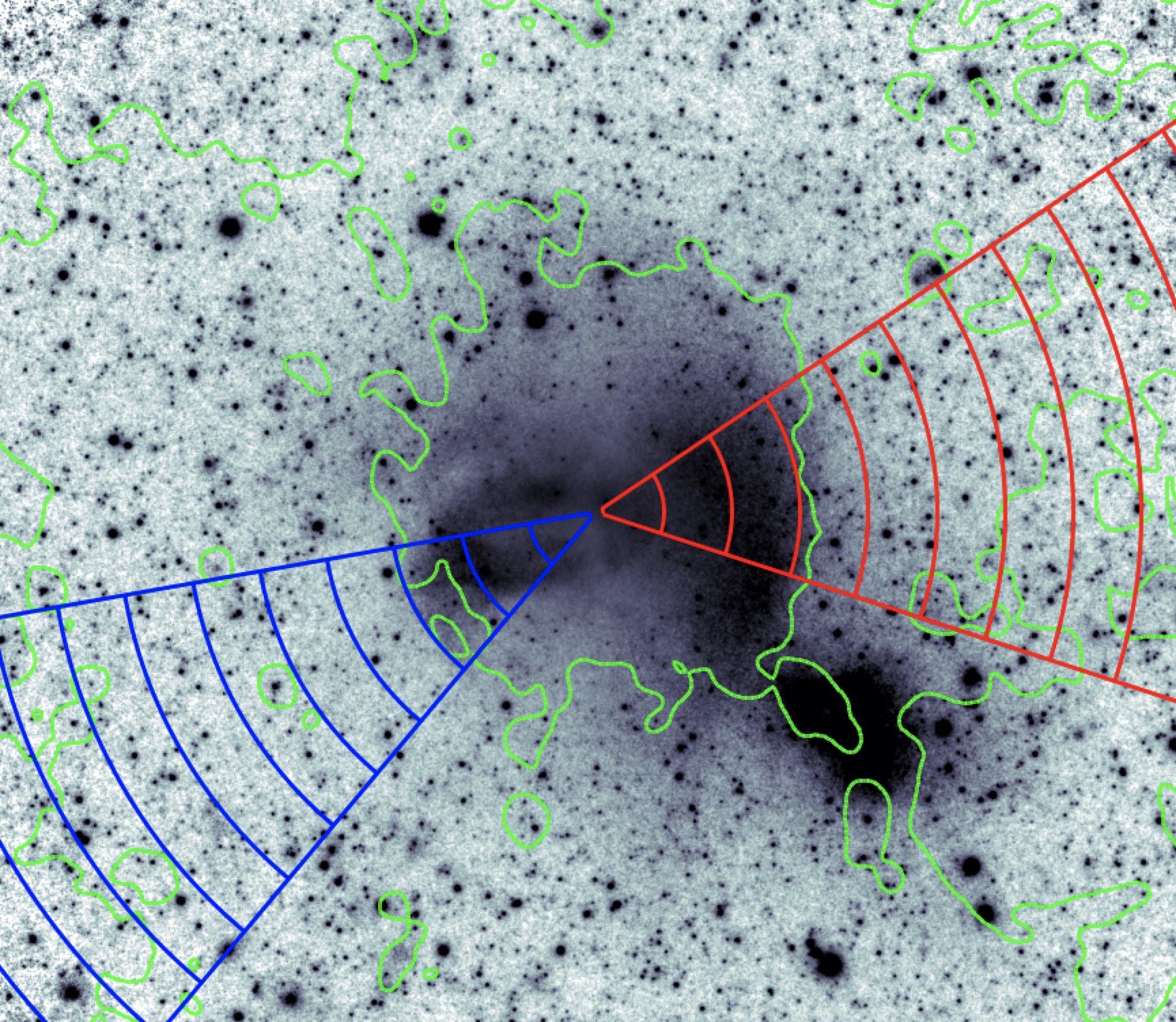}
\includegraphics[clip=t, width=0.49\linewidth]{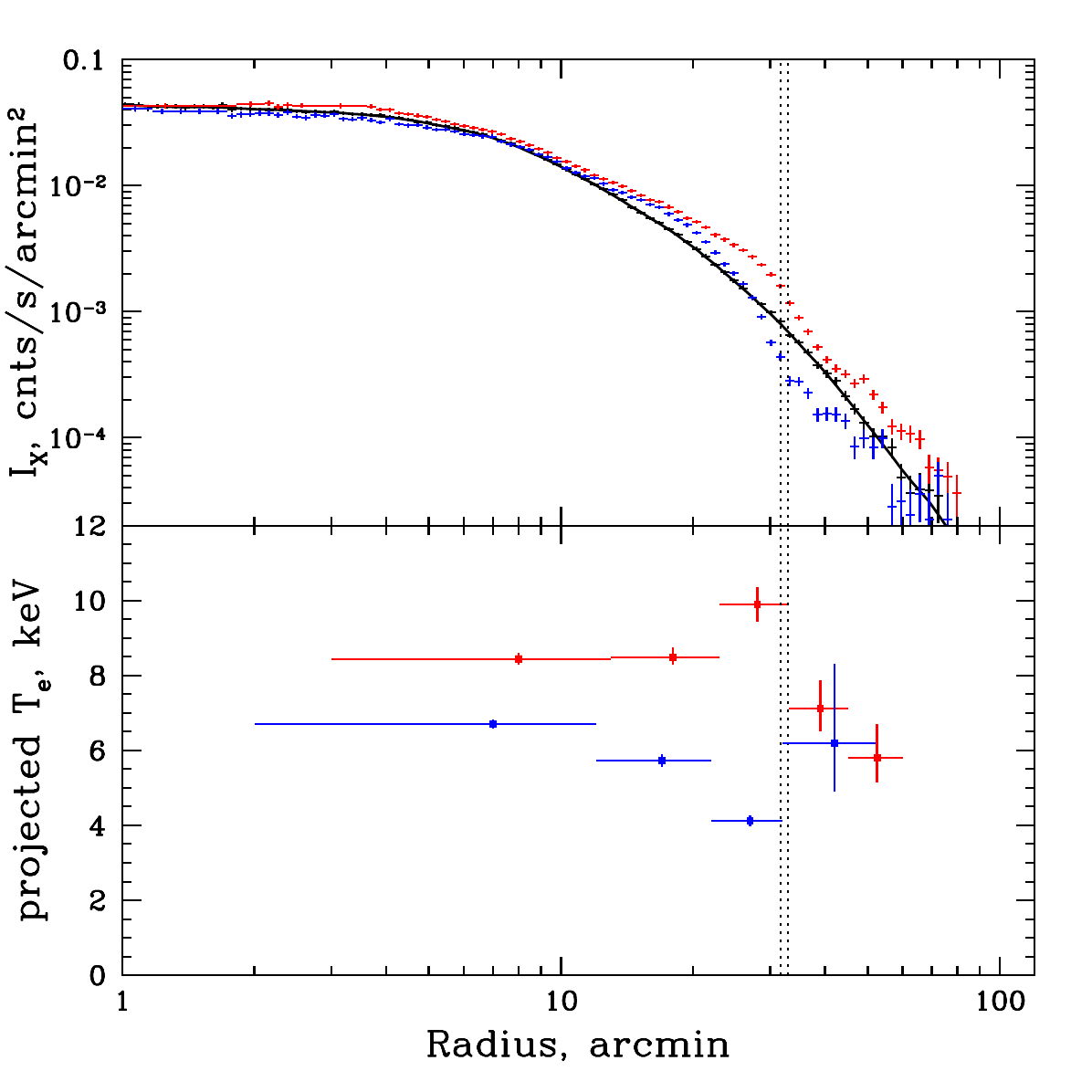}
}
\caption{{\bf Left:} Two wedges used for extraction of the X-ray surface brightness and the spectral analysis. The flattened X-ray image is used as a background. For convenience, 10' rings are shown. The green contours show the brightness of the WSRT radio image at 352~MHz  \citep{2011MNRAS.412....2B}. The inner contours approximately correspond to the boundary of the Coma radio halo {\bf Right:} Surface brightness and projected temperature profiles to the West (in red) and to the East (in blue) of the Coma center. To highlight surface brightness gradients, the top panel shows also the azimuthally averaged profile $I_X$ (the same as in Fig.~\ref{fig:coma_radial}) in black. To the West, sharp surface brightness discontinuity at $R\simeq 33$ arcmin coincides with the temperature decline. Thus, our observations confirm the presence of a shock wave in the West with the Mach number of $M \simeq 1.5$. The dotted line marks the position of the shock. To the East, we observe that the denser region is  colder than the more rarefied region, thus making the Eastern feature the cold front. The dotted lines show the positions of the X-ray surface brightness jumps. }
\label{fig:edges}
\end{figure*}

The quality of the present data set allows one to robustly classify the most prominent features seen in Fig.~\ref{fig:coma_flt}. In Figure~\ref{fig:edges}, we show surface brightness and (projected) temperature  profiles from two sectors, towards the West and the East of the cluster core with the position angles of (-18$^\circ$,40$^\circ$) and (190$^\circ$,230$^\circ$) respectively. 
Based on the observed radial profiles, we classify the West surface brightness edge as a shock front  with the Mach number of $\sim 1.5$, in agreement with the previous finding of \cite{2016PASJ...68S..20U, 2013A&A...554A.140P}. 

Although the edge feature to the East of the Coma core is interpreted as a shock in \cite{2013A&A...554A.140P}, the eROSITA temperature profile clearly shows that the denser region is actually colder than the less dense region, identifying this edge as a contact discontinuity. This does not exclude a possibility of a weak shock ahead of the cold front (farther to the East). A strong shock can be excluded since it would show up prominently in X-ray images. However, a weak (perhaps, transient) bow shock driven by a sloshing gas can not be entirely excluded by the current analysis. We defer the detailed discussion of this feature for subsequent studies. 

\section{X-ray and SZ images}

\begin{figure*}
\centering
\includegraphics[angle=0,trim= 0mm 0cm 0mm 0cm,width=1.6\columnwidth]{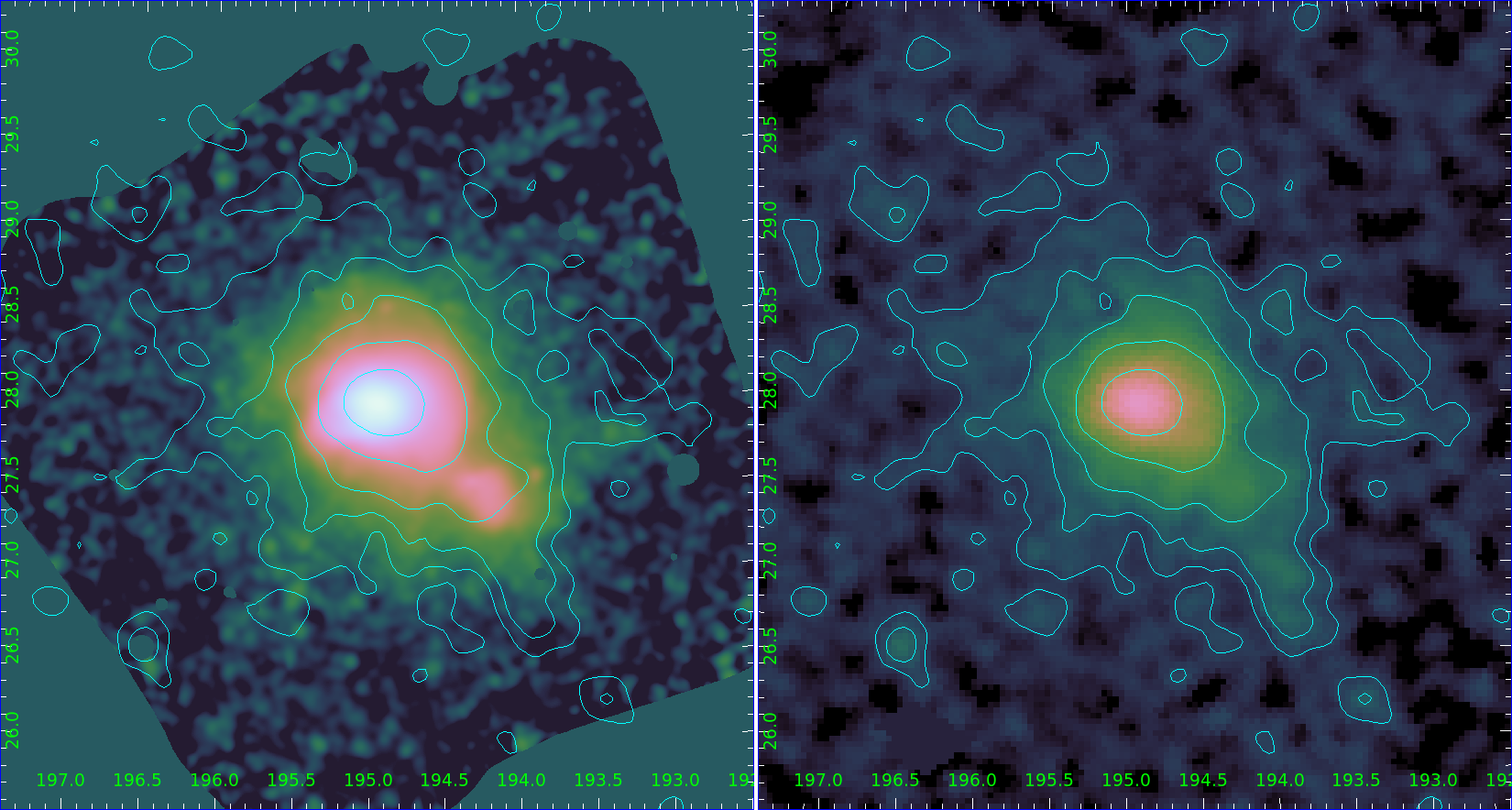}
\caption{0.4-2 keV X-ray and SZ images of the Coma cluster. Compact sources have been excised from the X-ray image before smoothing with the broad Gaussian ($\sigma=200"$). The size of the smoothing kernel was chosen to approximately match the angular resolution of the \textit{Planck} Y-map. Contours from SZ image are shown in both panels. Many non-axisymmetric features in these two images match suggesting that they are due to the presence of hot gas.}
\label{fig:coma_xsz}
\end{figure*}

Due to different dependencies of the X-ray and SZ signals on the gas density and temperature, it is possible to use them in order to constrain the gas thermodynamic properties or variations of these properties \citep[see, e.g.][]{2018A&A...614A...7G,2016MNRAS.463.1057C}. In particular, \cite{2020MNRAS.497.3204M} combined the \textit{XMM-Newton} and \textit{Planck} data on the Coma cluster in a number of wedges. With the extended spatial coverage provided by eROSITA, it is now possible to further advance the joint analysis of the SZ and X-ray data. To this end, we constructed a 2D map, which characterizes the density-weighted projected gas temperature distribution. When doing so, we made two simplifying assumptions.

First, we assumed that the X-ray emissivity in the 0.4-2 keV band is largely independent of the gas temperature, which a good approximation for hot clusters (see Appendix \ref{app:epsilon}), i.e. the X-ray surface brightness
\begin{eqnarray}
I_X = \int n_e^2 \epsilon_c(T) dl\approx \epsilon_c\left (\left \langle T \right \rangle \right ) \int n_e^2 dl,
\end{eqnarray}
where $\epsilon_c(T)\approx {\rm const}$ is known. The second assumption is that, to the first approximation, the surface brightness distribution (and the underlying 3D distributions of density) can be described by the $\beta$-model as shown above. 

\subsection{2D maps of the SZ and X-ray data}
\label{sec:xsz}
The X-ray and SZ images of the Coma cluster ($\sim 4.4\times 4.4$ degrees) are shown in Fig.~\ref{fig:coma_xsz}. The X-ray image is the same as shown in Fig.~\ref{fig:coma_ximage} with compact sources excised and smoothed with a Gaussian filter ($\sigma=200"$). The SZ image shown is largely similar to the one used in \cite{2013A&A...554A.140P}.  The contours in both images come from the SZ data; the lowest contour corresponds to the Compton parameter $y=2.5\,10^{-6}$. Each subsequent contour is factor of 2 higher. Here
\begin{eqnarray}
y = \frac{\sigma_T}{m_ec^2}\int kT n_e dl\equiv k\tilde{T} \frac{\sigma_T}{m_ec^2}\int n_e dl,
\end{eqnarray}
where $\tilde{T}$ is the electron-density-weighted temperature $\displaystyle \tilde{T}=\int T n_e dl/\int n_e dl$.  

 The close correspondence of the faint diffuse structures in X-rays and SZ images, which extend beyond $R_{500}$ is clear, including a well known extension in the direction of NGC~4839, but also fainter extensions to the West from the core and in the SE direction, which can already be recognized in the flattened X-ray image, albeit barely (Fig.~\ref{fig:coma_flt}). This correspondence suggests that these extensions are due to diffuse gas rather than a combination of faint unresolved compact X-ray sources. 

While the appearance of the X-ray and SZ images is largely similar, the quality of the data in the central region is sufficient to make a point by point comparison on spatial scales large enough so that the limited angular resolution of the \textit{Planck} data is not a major issue.  

The electron-density-weighted temperature $\tilde{T}$ can be expressed through the ratio of the two images as 
\begin{eqnarray}
k\tilde{T}\approx \frac{y}{I_X}\frac{m_ec^2}{\sigma_T}\frac{1}{4\pi}\frac{1}{(1+z)^2}\epsilon_c \left (\frac{n_H}{n_e}\right )\frac{\int n_e^2 dz}{\int n_e dz},
\label{eq:kt_from_szx}
\end{eqnarray}
where $I_X$ is the surface brightness in $\rm counts\,s^{-1}\,sr^{-1}$, $\epsilon_c$ is the eROSITA  counts "production rate" per unit emission measure for a plasma with a given temperature. For eROSITA $\epsilon_c(T)\approx {\rm const} \approx 50\times 10^{-14} {\rm counts\,s^{-1}\,cm^5}$ (see Appendix~\ref{app:epsilon}); $\left (\frac{n_H}{n_e}\right )\approx 0.83$ for the ICM. The last term can be evaluated only under assumption of a known 3D gas density distribution. For a beta model, this term reduces to 
\begin{eqnarray}
\frac{\int n_e^2 dz}{\int n_e dz}=\frac{n_{e,0}}{\left [1+\left( \frac{r}{r_c}\right )^2 \right ]^{3/2\beta}},
\end{eqnarray}
where $n_{e,0}\sim 4\,10^{-3}\,{\rm cm^{-3}}$ is taken from deprojected spectral analysis. Since eROSITA images have higher spatial resolution than the \textit{Planck} ones, the above correction factor was first applied to the eROSITA  0.4-2 keV image, which was then smoothed with a broad Gaussian ($\sigma\approx 250''$) to match (approximately) the angular resolution of \textit{Planck}. Compact sources have been subtracted from the image prior to smoothing. To avoid excessive noise at the edges of the map, the ratio was computed only over the area, where the surface brightness of the Coma emission is larger than the $\sim$25\% of the surface brightness due to the nominal background. 

\begin{figure}
\centering
\includegraphics[angle=0,trim= 0mm 5cm 0mm 3cm,width=0.95\columnwidth]{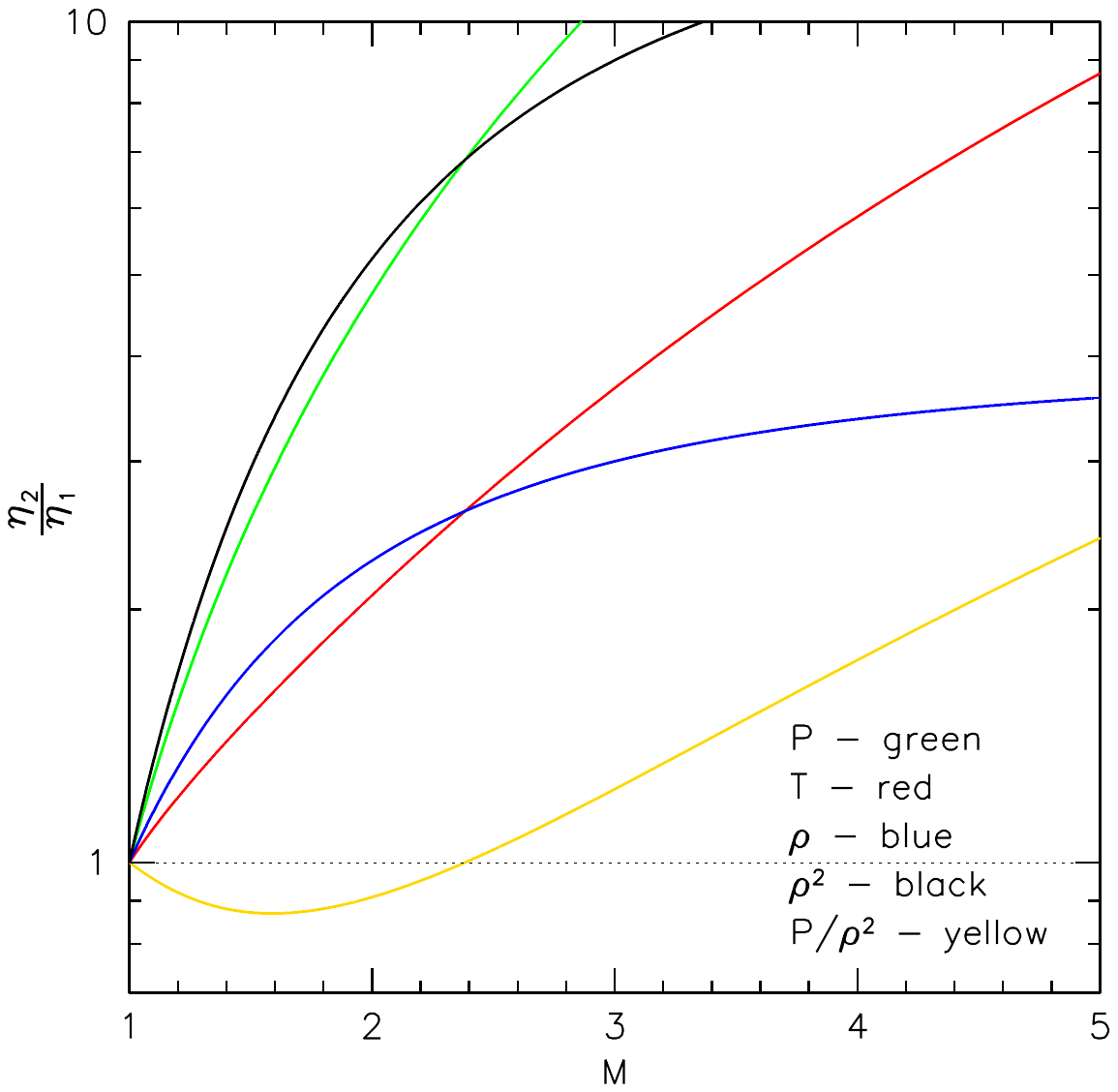}
\caption{Comparison of the "X-ray" and "Y" jumps for a shock in a gas with adiabatic index $\gamma=5/3$ as a function of the shock Mach number $M$. Rankine-Hugoniot conditions provide jumps in thermodynamic properties ($\rho$, $T$ or $P$). The ratio of $P/\rho^2$ approximately characterizes the ratio of the jumps in effective volume "emissivities" for X-rays and tSZ upstream and downstream of the shock. For $M\lesssim 2.4$, the jump in $Y$ is smaller than in X-ray volume emissivities, implying that the shock appears slightly less prominent in $Y$ than in the X-ray surface brightness. The minimum of the ratio (ignoring the temperature dependence of X-ray emissivity) is at $M\approx1.6$ and is equal to 0.87.}
\label{fig:szx_hugoniot}
\end{figure}

\begin{figure}
\centering
\includegraphics[angle=0,trim= 0mm 0cm 0mm 0cm,width=0.95\columnwidth]{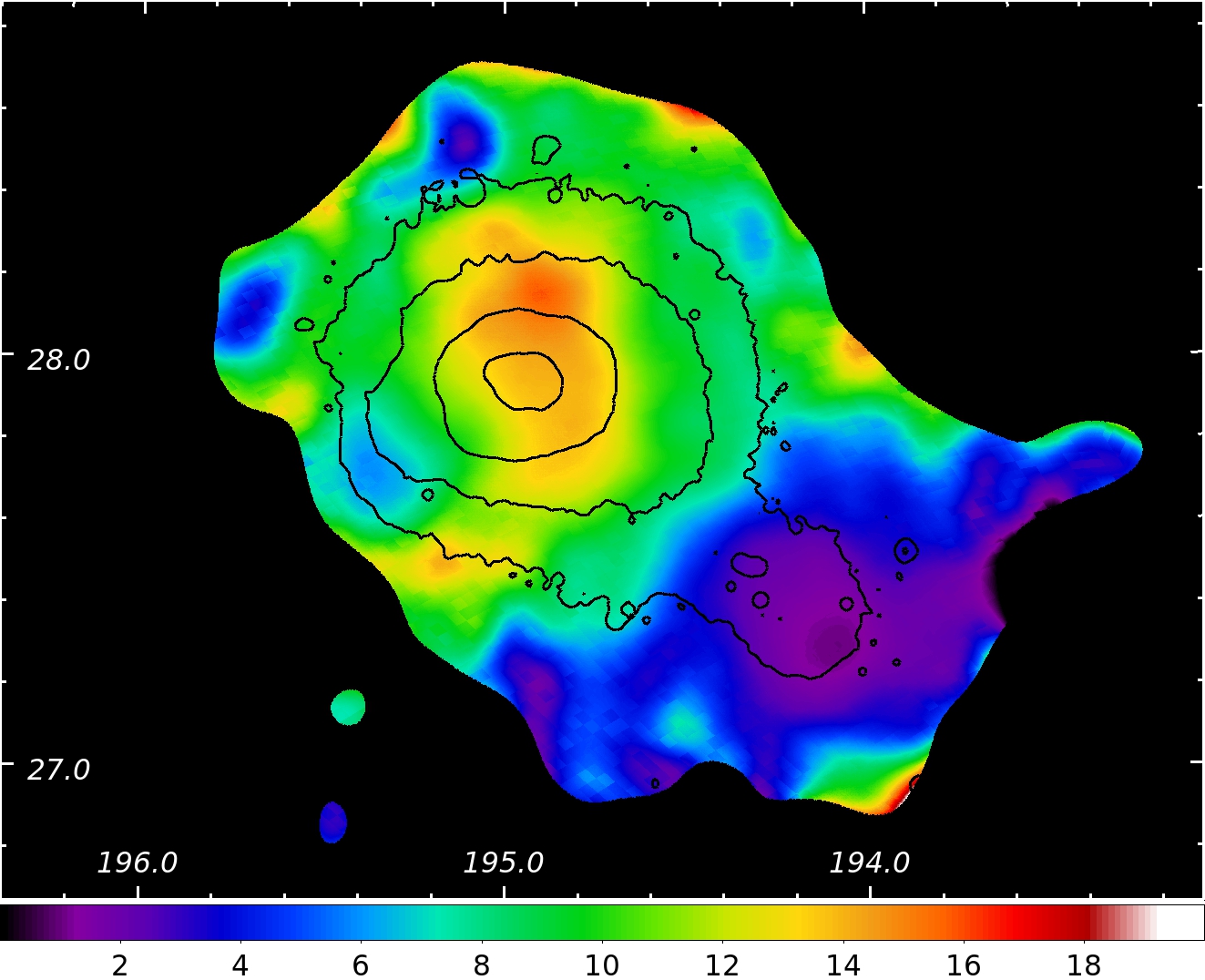}
\caption{Ratio of SZ and X-ray images of the Coma cluster, converted to the weighted electron temperature. Contours show the X-ray surface brightness. Only the inner part of the studied region, where the signal is strong enough to evaluate the ratio of two images is shown. There are three clear features in the derived temperature distribution - a cool region at the position of the NGC~4839 group, a (marginally) hotter region to the North from the core and another cool region near the contact discontinuity to the East from the core.}
\label{fig:coma_x_divsz}
\end{figure}

Overall, the recovered temperature map agrees well with the X-ray measurements of the projected temperature, which show $\sim 8-10$~keV in the core \citep[e.g., a temperature map in][]{2013Sci...341.1365S,Lyskova2019}. In the eROSITA+\textit{Planck} map, this temperature range corresponds to green color. To the North from the core, the temperature increases to $\sim 15$~keV. This is slightly higher than typically found in X-ray data, although a tentative presence of a "hot spot" on the North  was reported earlier \citep[e.g.][]{1999ApJ...513..690D,1999ApJ...527...80W}. Also, the central part of the map is in reasonable agreement with the recent  \textit{XMM-Newton}+\textit{Planck} temperature map by \cite{2020MNRAS.497.3204M}.
There are other prominent features in this image. First, there a "cold" spot towards the location of NGC~4839, well consistent with the presence of dense and relatively cold gas in this group \citep[see, e.g., a temperature map in][]{Lyskova2019,2020MNRAS.497.3204M}. 
Another "cold" spot is to the SE from the core, which was seen in the \textit{ASCA}, \textit{Chandra} and \textit{XMM-Newton} maps \citep[e.g.][]{1999ApJ...513..690D,1999ApJ...527...80W,Neumann.et.al.2003,2013Sci...341.1365S,2020MNRAS.497.3204M} and is tentatively associated with the infall of a galaxy group \citep[e.g.][]{1997ApJ...474L...7V}. 

We note in passing that the manipulation with X-ray and SZ images described above is expected to produce a robust and meaningful characterization of the gas temperature only if the adopted 3D model (a beta-model in our case) is correct. Presence of the NGC~4839 group obviously violates this assumption. Another structure affected by the same problem is the sharp discontinuity to the West from the core, which is a Mach $\sim 1.5$ shock (see Fig.~\ref{fig:coma_flt}) that does not show up prominently in this image. One possible reason for that might be a long electron-ion equilibration time on the downstream side of the shock, but 
for $M\sim1.5$ this effect is expected to be small. There are, however, two other reasons why the "cold" spots appear more clearly in 2D images as discussed below.


If we neglect the temperature dependence of $\epsilon_c$, the amplitudes of the X-ray and SZ signals at the shock front can be immediately compared as the ratio of the electron pressure and density squared jumps for the standard Rankine-Hugoniot conditions (assuming that electron and ion temperatures are equal). Namely,
\begin{eqnarray}
YX_{jump}\equiv\frac{P_d/P_u}{\rho_d^2/\rho_u^2}=\frac{(2+M^2(\gamma-1))^2(1+\gamma(2M^2-1))}{M^4(1+\gamma)^3},
\end{eqnarray}
where $M$ is the shock Mach number, and the subscripts $u$ and $d$ correspond to the upstream and downstream sides, respectively. The minimum of $YX_{jump}$ (see Fig.~\ref{fig:szx_hugoniot}) is at 
\begin{eqnarray}
M_{min}=\left ( \frac{\gamma+\sqrt{\gamma(\gamma(5-2\gamma)-2)}}{\gamma(\gamma-1)}  \right )^{1/2}.
\end{eqnarray}
For $\gamma=5/3$, $M_{min}\approx 1.6$ and at this Mach number $YX_{jump,min}\approx 0.87$; while at $M\approx 2.4$ the jump amplitudes are equal.

Now consider the behavior of the ratio of SZ and X-ray images (see Eq.~\ref{eq:kt_from_szx}) when thermodynamic properties of a small lump (of otherwise isothermal)  gas with  along the line of sight are modified, so that electron pressure changes by a factor $1+\eta_y$, while the density squared changes by another factor $1+\eta_x$, where $\eta_y\ll 1$ and $\eta_y\ll 1$. The smallness of $\eta_y$ and $\eta_y$ is needed here only to illustrate more clearly their impact on the derived temperature (by making Taylor expansion).  In this case, the following variation of the gas temperature, derived from Eq.~\ref{eq:kt_from_szx} is expected
\begin{eqnarray}
\tilde{T} \propto \frac{(1+\eta_y)y+Y}{(1+\eta_x)x+X}\approx\frac{y+Y}{x+X} \left (1+ \eta_y\frac{y}{y+Y} - \eta_x\frac{x}{x+X}\right )\propto \\
T_0 \left (1+ \eta_y\frac{y}{y+Y} - \eta_x\frac{x}{x+X}\right ),
\label{eq:tmod}
\end{eqnarray}
where $y,x$ are the contributions of this gas lump to integrated SZ and X-ray signals, respectively, while the capital $Y,X$ stay for the contributions due to remaining gas. $T_0=(y+Y)/(x+X)$ is the temperature of unperturbed gas. For isobaric perturbations, which can give rise to cold spots, $n_y=0$ and the change of temperature is always negative. For shocks, both $n_y$ and $n_x$ are positive and are almost equal, unless $M\gg 2.4$ as shown in Fig.~\ref{fig:szx_hugoniot}. Since they enter Eq.~\ref{eq:tmod} with different signs, the sign of the temperature change depends on the relation between $\frac{y}{y+Y}$ and $\frac{x}{x+X}$. In the case of the shock to the West from the core, these terms partly compensate each other and the resulting temperature does not change strongly. 
Similarly, some of the temperature variations seen in Fig.~\ref{fig:coma_x_divsz}, especially at outer regions, might also be induced by the very substantial deviations of the density distribution from the spherically-symmetric beta-model.

The above effects can be mitigated by adopting a more appropriate 3D model of the gas distribution, for example, a parametric model with a jump \citep[e.g.][]{2013A&A...554A.140P} or doing independent deprojection of X-ray and SZ data in wedges \citep[e.g.][]{2020MNRAS.497.3204M}, assuming that within each wedge all thermodynamic properties depend only on the radius. For the W-shock, the comparison of the eROSITA and \textit{Planck} data in a wedge yields consistent results (to be reported in a forthcoming publication).

\section{Merger signatures}
Massive galaxy clusters are dynamic systems that continuously accrete smaller structures. In the Coma cluster, the signatures of mergers have long been identified/suggested in the optical, radio or X-ray bands \citep[see, e.g.][and references therein]{2011MNRAS.412....2B,2020arXiv201108856B,Lyskova2019,2020A&A...634A..30M}.  Apart from the NGC~4839 group, these include the structures (shocks and contact discontinuities) discussed above. The famous radio relic has long been discussed in the association with a shock. Initially, it was suggested that the relic is located at the position of an accretion shock \citep{1998A&A...332..395E}, namely, where the infalling gas decelerates and joins the Coma ICM. An alternative is a scenario, where the shock is driven "from inside", i.e. by the merger with a smaller subcluster, which enters the main cluster from the opposite side and drives the shock which propagates outwards and takes over the "nominal" accretion shock. Here we discuss current observations in the light of this scenario.  

\subsection{Broad brush merger scenario}
Firstly, we outline the broad brush scenario, originally mentioned in \cite{Burns.et.al.1994, 1996A&A...311...95B} and discussed in detail in \cite{Lyskova2019, Sheardown2019, 2019MNRAS.488.5259Z}. In this scenario, the NGC~4839 group was initially moving from the NE along the general direction of the Coma-A1367 filament. It passed the pericenter for the first time some 1-1.5 Gyr ago, has already reached the apocenter, and is now falling back into the Coma center. This scenario was devised, in particular, to explain the apparent bent shape of the X-ray tail of NGC~4839. An immediate consequence of this scenario is the prediction of a "runaway" shock, which had been driven by NGC~4839 during its first infall, but has separated from the group after the pericenter passage and at present continues to propagate in the general SW direction. The decreasing density of the main cluster with radius helps this shock wave in keeping its strength (in terms of the Mach number) over a large traveled distance, even though the NGC~4839 group is already moving in the opposite direction. In this scenario the radio relic marks the position of this runaway shock and fits into a generic scenario of merger shocks \citep[e.g.][]{1997MNRAS.290..577R}. The subtlety here is whether this runaway shock has already taken over the accretion shock or not \citep[e.g.][]{2010MNRAS.408..199B,2020MNRAS.494.4539Z}. We deffer this question for the subsequent publication with the eROSITA results on the outer regions of the Coma cluster.

\subsection{Bridge between NGC~4839 and the main cluster}
One can argue that while crossing the pericenter and reversing direction at the apocenter, the NGC~4839 could leave a long tail of stripped gas tracing its trajectory through the main cluster, as long as this trace is not distorted/dissolved by the gas motions. We speculate that in the flattened X-ray image (Fig.~\ref{fig:coma_flt}) one can see a faint and curved sub-structure that can be a trace of this gas (see also Fig.~\ref{fig:coma_flt_label3} below, where this feature is labeled as "Bridge"). Unambiguous identification  of this feature as the stripped NGC~4839 gas is not robust, but a combination of accurate temperature and abundance measurements in X-rays and morphology of faint radio emission might help to do this.

\subsection{Sharp Western edge as a secondary shock}
Apart from the NGC~4839, the next most prominent feature seen in the eROSITA X-ray images (Figs.~\ref{fig:coma_ximage},\ref{fig:coma_central},\ref{fig:coma_flt}) is a long and sharp edge to the West from the Coma core, which spans more than 2 Mpc from North to South with its curvature changing from concave (closer to the Coma center) to convex at the farthest end on the South. The brightest part of this structure has already been recognized in the \textit{ROSAT},  \textit{XMM-Newton},  \textit{Planck} images \citep[e.g.][]{Neumann.et.al.2003,2013ApJ...775....4S,2013A&A...554A.140P,2020MNRAS.497.3204M} and reported to be a shock, although there was no consensus on the origin of this shock. Spectral analysis of eROSITA data (see \S\ref{sec:xsubstructure}) unambiguously identifies this structure as a shock. The brighter (closer to the Coma center) side is hotter, so this is the downstream side of the shock. 

Is this a classic accretion shock, i.e. the infall of the Intergalactic Medium (IGM), which is about to join the Coma cluster? There are two difficulties with this scenario. First, this shock is only $1\,$~Mpc (i.e. $\lesssim R_{500}$), from the Coma center, what is unlikely, unless there is a dense filament in this direction. Second, the diffuse emission can be seen on the upstream side of the shock, i.e. farther away from the Coma, which would require pre-heating the upstream gas to the temperature $\sim 5$~keV.

Is this a merger (or a bow) shock driven by the NGC~4839 group\footnote{Here we make a distinction between the bow shock which is currently driven by a moving group during its accelerated motion towards the center of the main cluster and the merger shock, which runs away from the moving group when it decelerates after passing through the pericenter \citep[see, e.g.][]{2019MNRAS.488.5259Z}}?
The bow shock hypothesis clearly contradicts the morphology of the front. Indeed, the bow shock would encompass the moving group and the hot (downstream) gas would be closer to the group. The observed geometry is just the opposite. The same is true for the merger shock scenario since in this case, the shock should be farther away from the Coma cluster (in the direction of the initial infall) than the group. For instance, it could be located at the position of the radio relic. 

With the above scenarios having been discarded, we are left with two other possibilities: i) there are two (or more) mergers happening now in Coma or ii) only one important merger (with the NGC~4839 group) is responsible for the bulk of the X-ray substructure. 

A plausible scenario, which predicts that two shocks (one far away from the core and another one close to the core) are produced by a single merger is discussed in \cite{2021MNRAS.501.1038Z}. Namely, during the first passage through the core, the smaller subcluster drives the bow/merger shock, which displaces the gas in the main cluster core. While the merger shock continues to propagate towards cluster outskirts, this gas eventually falls back to the core and settles (eventually) to hydrostatic equilibrium. While this is largely a scenario that is often invoked to explain gas sloshing in the core, the internal waves and global oscillations of the gas are not the only outcomes of this process. In addition, a "mini accretion shock" can be formed, when the primary merger shock is already far away from the center. 

\begin{figure*}
\centering
\includegraphics[angle=0,trim= 0mm 10mm 0mm 10mm,width=1\columnwidth]{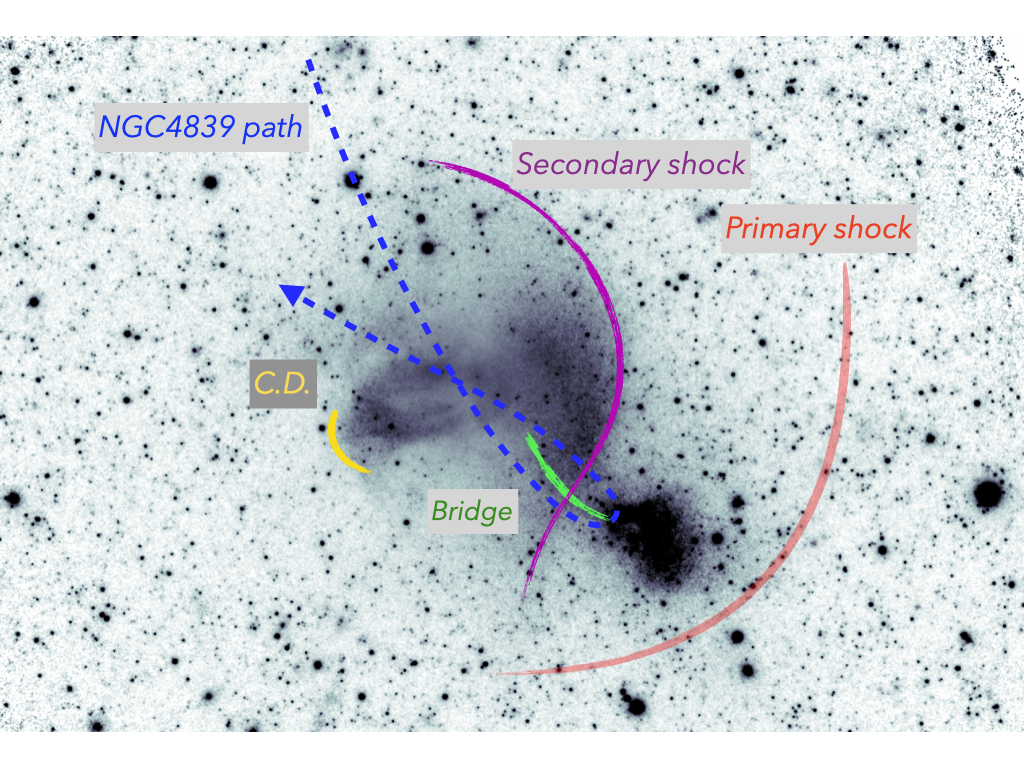}  
\includegraphics[angle=0,trim= 0mm 5mm 0mm 0cm,width=0.86\columnwidth]{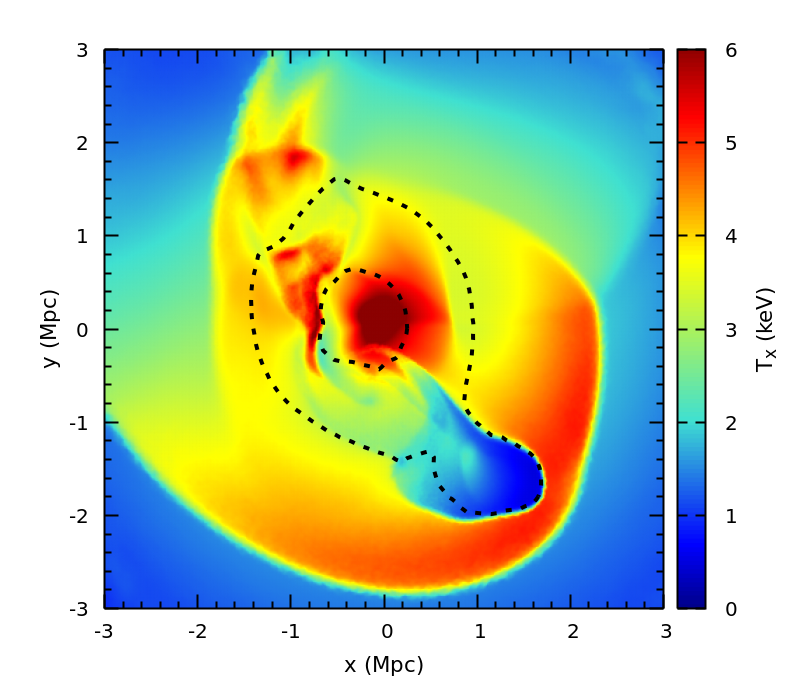} 
\caption{{\bf Left:} Flattened image of the Coma cluster field with labels schematically marking some of the features associated with the merger with the NGC4839 group. The blue dashed line is the suggested trajectory of the group \citep{Lyskova2019,Sheardown2019,2019MNRAS.488.5259Z}, which enters the Coma cluster from NE, and is currently close to the apocenter. The presumed position of the primary shock is shown with the red curve so that it goes through the radio relic SW from the NGC4839 group. The purple curve marks the secondary shock caused by settling the displaced ICM back to the hydrostatic equilibrium. This is the most salient feature directly seen in the image as the surface brightness edge. The green line shows the fain X-ray "bridge" connecting NGC4839 and the main cluster, which is a possible trace of the group through the Coma cluster. {\bf Right:} Projected temperature map from simulations \citep[][see their fig.8 for the original version of the figure]{2021MNRAS.501.1038Z}. 
The simulations qualitatively reproduce the morphology of the cold gas and the positions of the primary and secondary shocks. In the simulations, the inner region is a complex mixture of shocks and contact discontinuities. It is, therefore, not surprising that real observations have both types of structures.}
\label{fig:coma_flt_label3}
\end{figure*}

Examples of such shocks are given in the numerical simulations of \cite{2021MNRAS.501.1038Z}, which are specifically focused on this issue (see the right panel in Fig.~\ref{fig:coma_flt_label3}). One can also identify similar structures in publicly available generic simulations of cluster mergers. For instance, in the library of cluster mergers by \cite{2011ApJ...728...54Z}, the secondary shocks are particularly visual for the mass ratio $R1:10$, the impact parameter $b=500\,{\rm kpc}$ at $t\sim3$~Gyr (\verb#http://gcmc.hub.yt/fiducial/1to10_b0.5/index.html#). Yet another example, is the idealized simulation of a plane shock traversing the core of the cluster in \cite{2003ApJ...590..225C} (see their section 4) when both sloshing motions and shocks are formed. In the double shock scenario, both shocks are the natural outcome of a single merger and no extra perturbations are needed. One has to observe the cluster at a particular moment of the merger, namely when the sub-cluster is close to the apocenter during its first passage.

Apart from the merger and mini-accretion shocks, the merger should generate sloshing motions, which naturally generate contact discontinuities \citep[a.k.a. cold fronts, see][and reference therein]{2007PhR...443....1M}. It is therefore not surprising that some of the sharp edges in the Coma image correspond to contact discontinuities. The most prominent one is the sharp edge to the East from the core (labeled as "C.D." in Fig.~\ref{fig:coma_flt_label3}), for which the spectral analysis (see \S\ref{sec:xsubstructure}) shows that the brighter/dense side is cooler, as expected for a genuine contact discontinuity. 

In the above discussion, we are trying to attribute most of the morphological features observed in the X-ray band, in particular, the W-shock, to the NGC4839 group. However, optical and X-ray data suggest additional ongoing accretion/mergers in the Coma cluster. One example is a chain of galaxies in the NGC4911-NGC4921 group \citep[e.g.][]{1997ApJ...474L...7V,2013ApJ...766..107A} and another is the presence of two D galaxies NGC4889 and NGC4874 in the core of the cluster \citep[e.g.][]{1996A&A...311...95B,Adami.et.al.2005}. The contribution of these mergers to the formation of structures visible in X-ray images, e.g. of the W-shock, can not be excluded apriori. However, we believe that NGC4839 is still a primary driver of this shock. Indeed, the NGC 4911-NGC 4921 group is less massive \citep[e.g.][]{Adami.et.al.2005}  and appears to be just approaching the central part of Coma. Given the extent of the W-shock, it appears unlikely that this group could already make such an impact on the Coma gas. On the other hand, NGC4889 and NGC4874 are located close to the geometrical center of the main cluster, have a small line-of-sight velocity difference of $\sim 720\, \rm km\,s^{-1}$, and do not currently possess very massive cool gaseous atmospheres, characteristic for a group not yet stripped by the ram pressure. Such an atmosphere would act as a working surface, which drives a shock wave. Therefore, if one of the two galaxies is responsible for the W-shock \citep[see][for the discussion of various scenarios]{1996A&A...311...95B,Adami.et.al.2005} then it should have had the atmosphere less than $\sim$400~Myr ago (an upper limit based on the distance to the W shock from the center ($\sim 33'\sim 930\,{\rm kpc}$ and the current Mach number of the shock). The group also had to move fast at that time in order to drive the shock. This scenario would then require fine-tuning of the merger geometry and/or a very efficient dynamic friction to explain the current positions and velocities of NGC4889 and NGC4874. At the same time, we do not see any fundamental problems with a scenario when the cool/dense atmosphere(s) was lost a few Gyr ago and two galaxies are now spiraling into the core.   Given these arguments, it seems that the scenario involving NGC4839 is more natural.

\subsection{Coma radio halo}
The Coma radio halo has been intensively studied over several decades \citep[see, e.g.,][and references therein]{2011MNRAS.412....2B,2020arXiv201108856B}.
If the merger scenario and our identification of various features in the X-ray map are correct, this might have interesting implications for the formation of the radio halo in the Coma core on the downstream side of the  Western shock. Since in this scenario the Western shock is the "secondary shock" (or "mini-accretion shock"), the gas upstream of the shock has already passed through the primary shock soon after the NGC4839 group passage through the core about a Gyr ago. Let us assume that the primary shock was strong enough to accelerate (or re-accelerate) particles that can potentially emit synchrotron radiation \citep[see, e.g.][for a review]{2019SSRv..215...14B}. Can these particles survive for a Gyr and eventually been compressed (and re-accelerated) by the secondary shock? For a power law distribution of isotropic relativistic electrons in a randomly oriented magnetic field, the Lorentz factor $\gamma_{max}$ that provide the largest contribution to the flux emitted at frequency $\nu$ is
\begin{eqnarray}
\gamma_{max}=\zeta_{max}\gamma_c,
\end{eqnarray}
where $\gamma_c$ is set by the condition
\begin{eqnarray}
\nu=\nu_c\equiv\frac{3}{4\pi}\frac{eB}{m_ec}\gamma^2_c.
\end{eqnarray}
Here, $B$, $e$, $m_e$ and $c$ are magnetic field, electron charge, electron mass and the speed of light, respectively.
$\zeta_{max}$ is the function of the electron spectrum slope, which is linked to the observed spectral index of synchrotron emission $\alpha$. For  $\alpha\sim2$, $\zeta_{max}\sim 0.57$ (see Appendix~\ref{app:gamax}). Moreover, the turbulent character of the cluster magnetic field may further reduce the value of $\zeta_{max}$.

The life time of a relativistic electron with the Lorentz factor $\gamma_{max}$  against synchrotron and Inverse Compton losses is 
\begin{eqnarray}
t_{cool}=-\frac{\gamma}{\dot{\gamma}}=\frac{1}{\zeta_{max}}\left(\frac{\nu}{\frac{3}{4\pi}\frac{eB}{m_ec}}\right)^{-1/2}\left ( \frac{4}{3}\frac{\sigma_T}{m_ec}\right )^{-1} \nonumber \\ \times \left ( \frac{B^2}{8\pi}+U_{CMB}\right )^{-1},
\end{eqnarray}
where $\sigma_T$ is the Thomson cross section, and $U_{CMB}$ is radiation field energy density (dominated by the Cosmic Microwave Background, CMB). At redshift $z\approx 0$, $U_{CMB}$ is equivalent to the magnetic field energy density $\frac{B^2}{8\pi}$ for $B\approx 3.25\,{\rm \mu G}$. The cooling time $t_{cool}$ reaches the maximum for $B\sim 2\,{\rm \mu G}$.  Conveniently, Faraday Rotation measurements suggest $B\sim 1-2 \mu G$ at the distance of $\sim 700\,{\rm kpc}$ from the Coma center \citep{2010A&A...513A..30B}. Therefore, the life time of electrons is not far from it maximum value (plugging $B=2\,{\rm \mu G}$ in the above expression) 
\begin{eqnarray}
t_{max}=1.3\,10^{8} \frac{1}{\zeta_{max}} \left ( \frac{\nu}{1.4~{\rm GHz}} \right )^{-1/2} \nonumber \\ \approx 1.7\,10^{8} \left ( \frac{\nu}{1.4~{\rm GHz}} \right )^{-1/2}\;{\rm yr}.
\end{eqnarray}

For 150 and 350 MHz, the cooling time is $5.3\,10^8$ and $3.5\,10^8$~yr, respectively.  While long enough, 
these time scales are still shorter than $\sim$~Gyr elapsed since the passage of the primary shock through the gas, which is now close to the secondary shock. However, downstream of a spherical merger shock (the primary shock), the compressed region is followed by a rarefaction, where the density falls below the initial density of a given gas parcel (see fig.5 in \citealt{2019MNRAS.488.5259Z}). This rarefaction is caused by the expansion of the gas that is pushed by the shock to larger radii.  This means that for a significant fraction of time between the passages through the primary and the secondary shocks, the electrons have lower Lorentz factors than just behind the primary shock, and the magnetic field is also lower. The radiative losses are therefore relatively small (except for the period right after the passage of the primary shock). The electrons then spend a long time in a "freezer", until the gas passes through the secondary shock (mini-accretion shock), when its density increases back to the initial state (at least approximately).  The electrons' Lorentz factor and the strength of the magnetic field increase accordingly. As a result, the synchrotron emissivity is boosted by a factor $F=C^{\frac{2}{3}p+1}$ \citep[e.g.][]{2005ApJ...627..733M}, where $p$ is the slope of the relativistic electrons energy spectrum and  $C$ is the volume compression factor. In terms of the spectral index $\alpha=(p-1)/2$ in the radio band, the boost factor is $C^{\frac{4\alpha}{3}+\frac{5}{3}}$. This expression is based on the assumption that the magnetic field is enhanced by a factor $C^{2/3}$.  For the shock with $M\sim 1.5$, the compression is $C\approx 1.7$. For $\alpha=2$, the corresponding boost of the radio flux purely due to adiabatic compression is $F\approx 10$. Unlike the "runway" merger shock, the mini-accretion shock does not feature a rarefaction region downstream and, therefore, the radio emission is controlled by radiative losses. Once again the question arises whether the short radiative times pose a problem?  The distance over which the secondary shock front can travel during $t_{max}$ is $l_{max}\sim c_s t_{max}\lesssim 180\,{\rm kpc}$ (for $\nu=1.4\,{\rm GHz}$), assuming that the shock moves with the sound speed\footnote{We note in passing that the velocity of a mini-accretion shock relative to the Coma center can be slightly lower than the sound speed.}. At $\nu=150\,{\rm MHz}$, this distance will be $\sim$500~kpc. This is smaller than the size (radius) of the radio halo of some 700~kpc (not by a huge factor, though), which used to be an argument against the single shock acceleration scenario. However, in the scenario, where secondary shocks are responsible for the compression and, possibly, reacceleration of electrons, this problem can be alleviated. The electron reacceleration scenario implies that there is a pool of non-thermal relativistic electrons produced by MHD shocks and turbulence due to the cluster merging history and fast moving galaxies in the cluster. Then the diffusive shock acceleration (DSA) mechanism would reproduce the slope of the pool electron spectrum in a shock downstream if the pool distribution had a power-law spectrum 
of index $p < \frac{C + 2}{C - 1}$. Otherwise, the spectrum in the shock downstream would have the standard test particle DSA index $p = \frac{C+2}{C -1}$ if the slope of the pool distribution has a steeper spectrum \citep[see e.g.][]{1987PhR...154....1B}. Therefore, the shock with $M\sim 1.5$, and the compression ratio $C\approx 1.7$ propagating in the cluster plasma regions without the prominent pre-existing non-thermal population  have $p=5.2$ and the synchrotron radio emission index $\alpha=2.1$. On the other hand, energetic particle  acceleration by intensive MHD turbulence and multiple shocks may produce a long-lived pool of relativistic particles  \citep[see e.g.][]{1987A&A...182...21S,2000A&A...362..886B,2001MNRAS.320..365B,2008SSRv..134..207P,2011MNRAS.412..817B,2019SSRv..215...16V}. Namely, a time dependent non-linear model of electron acceleration by the shock ensemble with long-wavelength MHD turbulence predicts the time asymptotic particle spectra with power-law  indexes $p \geq$ 3 providing synchrotron radio emission of indexes $\alpha \sim 1$ \citep[see Fig. 7 in][]{2019SSRv..215...14B} which is consistent with that is observed in the extended  regions of the cluster core \citep[see][and references therein]{2011MNRAS.412....2B}.        

Although the multiple shocks operate in different parts of the cluster core, they are, however, synchronized by the initial merger shock.  To some extent, these "coordinated" shocks could provide a similar effect as the re-acceleration by the turbulence in the core, which was invoked as an alternative mechanism to keep the radio halo alive \citep[see, e.g.][]{2001MNRAS.320..365B,2020arXiv201108856B}. Given the long-living perturbed state of the core predicted by simulations, a combination of both scenarios could be at work in the Coma core. We also note that the sloshing of the gas in the core, when big lumps of gas temporarily move away from the core and expand, essentially quenching the radio emission, could be another reason for the appearance of a sharp boundary of the halo, like observed to the East from the Coma center.

Is it possible to devise an observational test that can support or falsify the above scenario (or at least parts of it)? First of all, the spatial coincidence of the radio halo edge and the shock clearly shows that the compression (or compression+acceleration or re-acceleration) is happening right at the position of the W shock. As discussed above, a weak shock would not alter the slope of the particles’ distribution, unless the distribution is very steep. Therefore, if sensitive low-frequency observations (e.g., LOFAR, GMRT) could detect diffuse radio emission outside the W shock and compare slopes inside and outside the shock, the role of the preexisting population and the revival scenario can be clarified. The same slope and correct intensity ratio would support a pure compression scenario. Further inside the radio halo, the “volumetric” acceleration mechanisms and spectral aging might be important and their roles could be revealed by measuring the radial gradients of the spectral indices at low and high frequencies. In the absence of volumetric acceleration, the spectral aging (at high frequencies) should become more prominent towards the center, while the lowest frequency part might show an opposite trend, which reflects the high Mach number of the shock that went through the fluid element close to the center. Furthermore, in the outgoing shock scenario (see also the discussion in the previous subsection), the expansion downstream of the shock could contribute to steepening of the high-frequency end of the spectrum.

If the above scenario is correct, a typical post-merger configuration might include two radio-bright structures. The outer one (radio relic), located close to the position of the primary shock - is formed due to particle acceleration and/or compression at the shock, followed by adiabatic expansion and radiative losses at high frequencies (see, e.g., fig.7 in \citealt{2019MNRAS.488.5259Z}). The inner one, inside the mini-accretion shock, is a more diffuse and extended radio halo. These two radio-bright regions should be separated by a radio faint "ring", which corresponds to a rarefaction region between the mini-accretion shock and the primary shock.

The broad-brush single-shock scenario \citep[see, e.g.][]{2010arXiv1010.3660M} tailored to the Coma case, implies that the Western edge of the halo is a merger shock with a moderate Mach number that re-accelerates electrons, while the bulk of the halo radio emission is due to turbulent re-accelerations. The scenario outlined above keeps these elements, but adds two interesting ingredients. Firstly, it postulates that a genuine primary shock (with higher Mach number) must be present ahead of the Western (mini accretion) shock. This alleviates the “preparation” of electrons for the Western shock compression and, possibly, re-acceleration. Secondly, in this scenario,  the mini accretion shock has a complicated 3D shape \citep{2021MNRAS.501.1038Z} and might envelope the entire cluster core (at least in projection), rather than boosting radio emission only over one side of the cluster, as the primary shock does. Moreover, the complex MHD-flows associated with the secondary shocks would energize turbulence over the cluster volume similarly to the single-shock scenario. Indeed, turbulent re-acceleration has been considered in connection with low polarization of the halo, overall halo spectrum, and the volumetric correlation of X-ray-radio brightness \citep[see, e.g.][]{2001A&A...369..441G,2013A&A...558A..52B,2011MNRAS.412....2B}. Detailed simulations of the double-shock scenario (to be reported elsewhere) might help to clarify the roles of the two shocks and the turbulence in explaining these radio halo properties. 

\section{Conclusions}
Observations of the Coma cluster with \textit{SRG}/eROSITA revealed a rich substructure in the X-ray surface brightness distribution, extending to the cluster virial radius. Analyzing this substructure we came to the following conclusions:

\begin{itemize}
\item The overall morphology is consistent with the post-merger scenario described in \cite{Lyskova2019, Sheardown2019, 2019MNRAS.488.5259Z}, in which the NGC~4839 group has passed through the Coma cluster from the NE direction, reached the apocenter, and is now starting its second 
infall towards the main cluster. 

\item A faint X-ray bridge connecting the NGC~4839 with the main cluster provides convincing proof that the group has already crossed the Coma core.

\item X-ray spectral analysis robustly classifies some of the most prominent structures as shocks or contact discontinuities. In particular, the steepest gradient in the X-ray surface brightness $\sim$30-40$'$ to the West from the core appears to be a shock, while the steepest gradient to the East is a contact discontinuity.

\item In the post-merger scenario, the most prominent "edge" to the West from the core is interpreted as a "mini-accretion" (or "secondary") shock \citep[see the simulation in  ][]{2021MNRAS.501.1038Z}, associated with the infall of the gas displaced by the merger with the NGC~4839 group on its first passage, and settling of this gas back into hydrostatic equilibrium. This scenario envisages that the primary merger shock is currently located at the position of the radio relic.

\item The presence of two shocks leads to the following scenario of the radio halo formation in the Coma cluster. The gas encompassed by the secondary shock has already passed through two shocks, both the primary and the secondary. The primary shock, which presumably is the main source of particle acceleration, is followed by a rarefaction that lowers particles' Lorentz factors and extends their radiative lifetime. These particles begin radiating again only after crossing the secondary shock, which lacks the rarefaction. While the radiative lifetime of electrons emitting at hundreds of MHz - 1 GHz is not very long, compared to, say, sound crossing time of the halo, the "synchronization" of secondary shocks in the core by the initial merger leads to the appearance of the extended and steep spectrum halo.

\item The comparison of the X-ray and SZ (\textit{Planck}) images leads to a 2D gas-density-weighted temperature map, which reveals a number of hot and cold spots, although the detailed structure might be affected by the different angular resolution in the X-ray and microwave bands and imperfection of the cluster 3D model. We have shown that pure entropy perturbations (the pressure is not affected) are expected to be robustly revealed by such maps. For shocks with low Mach numbers ($M\lesssim 3$), the "weighted" temperature depends on the relative contributions of SZ and X-ray signals from the perturbed region to the total signals along a given line of sight.      
\end{itemize}

This study is intentionally restricted to features that are the most robust against remaining calibration uncertainties, in particular, the ones associated with the single-scattering stray light and careful calibration of the telescope effective area. The former is especially important for the outskirts of the cluster, i.e. $R_{200}$ and the outer baryonic boundary of the Coma cluster. The latter is crucial for accurate temperature determination. A more detailed analysis will be reported in subsequent publications.  

\section{Acknowledgments}
We thank our referee for useful comments that helped to improve the paper.

We are grateful to Congyao Zhang for providing a modified version of the figure based on simulations described in \cite{2021MNRAS.501.1038Z} and to Larry Rudnick for providing us with the 352-MHz FITS image of the Coma cluster published in \cite{2011MNRAS.412....2B}.

This work is based on observations with eROSITA telescope onboard SRG space observatory. The SRG observatory was built by Roskosmos in the interests of the Russian Academy of Sciences represented by its Space Research Institute (IKI) in the framework of the Russian Federal Space Program, with the participation of the Deutsches Zentrum für Luft- und Raumfahrt (DLR). The eROSITA X-ray telescope was built by a consortium of German Institutes led by MPE, and supported by DLR. The SRG spacecraft was designed, built, launched and is operated by the Lavochkin Association and its subcontractors. The science data are downlinked via the Deep Space Network Antennae in Bear Lakes, Ussurijsk, and Baikonur, funded by Roskosmos. The eROSITA data used in this work were converted to calibrated event lists using the eSASS software system developed by the German eROSITA Consortium and analysed using proprietary data reduction software developed by the Russian eROSITA Consortium.

EC, IK, NL, and  RS  acknowledge  partial  support  by  the  Russian  Science  Foundation  grant  19-12-00369. 

This research has made use of data obtained from XMMSL2, the Second \textit{XMM-Newton} Slew Survey Catalog, produced by members of the XMM SOC, the EPIC consortium, and using work carried out in the context of the EXTraS project ("Exploring the X-ray Transient and variable Sky", funded from the EU's Seventh Framework Programme under grant agreement no. 607452). 

This research made use of \texttt{Montage}. It is funded by the National Science Foundation under Grant Number ACI-1440620, and was previously funded by the National Aeronautics and Space Administration's Earth Science Technology Office, Computation Technologies Project, under Cooperative Agreement Number NCC5-626 between NASA and the California Institute of Technology.

Catalog manipulations have been performed using the \texttt{TOPCAT/STILTS} software \citep{2005ASPC..347...29T}.


\appendix
\section{Provisional background model}
\label{app:bg}
To correct for the instrument intrinsic background, Galactic foreground and astrophysical background in the Coma field, we have used a "provisional" background model.  We model the total background spectrum per unit solid angle $B_t$ as 
\begin{eqnarray}
B_t=B_i+B_G+B_{CXB}.
\end{eqnarray}
where $B_i$, $B_G$, $B_{CXB}$ are the intrinsic detector background, Galactic foreground (mostly diffuse emission), astrophysical background (mostly AGN), respectively. In the survey or scan mode, the vignetting can be assumed constant over the studied  region, which simplifies the treatment of the background. 
The advantage of the SRG observatory being at L2 point is the notable stability of the background. It implies that  the background measured during one observation can be used for the analysis of another observation several months earlier or later (at long as background variations at the per cent level are not crucial for the problem at hand).
Thus if only the total background rate matters, one can use any other field with similar Galactic hydrogen column density. However, in the Coma field, the sensitivity to compact sources, which have to be excised before evaluating the X-ray surface brightness, varies across the field. For that reason the remaining background has to be adjusted for the resolved fraction. To this end, we estimated the resolved part of the background by integrating the Log~N-Log~S curve taken from Luo+17. We have verified using several fields observed with eROSITA that we are getting consistent Log~N-Log~S and the conversion factors to our working energy bands is reasonable. With this approach, the evaluation of the background in a given energy band (and for a given flux level $s_{min}$ of resolved and excised sources) reduces to:
\begin{eqnarray}
B=B_t-\int_{s_{min}}^\infty \frac{dN}{ds}sds,
\end{eqnarray}
where $\frac{dN}{ds}$ is derived from Chandra's 0.5-2 keV Log~N-Log~S. The images used in this work were made in the 0.4-2 keV band. In this band, $B_t\sim 4.24\,10^{-4}\,{\rm counts\,s^{-1}\,arcmin^{-2}}$ per one (out of 7) eROSITA telescopes,  $B_i$ makes about a quarter of it.

\section{The model of the effective Point Spread Function}
\label{app:psf}
In-flight calibration of the \srg/eROSITA's effective Point Spread Function (PSF) is on-going. In particular, we performed a stacking of the properly-aligned images of known point sources observed during the course of the all-sky survey. The selection of sources is based on their extragalactic position and X-ray brightness, which should not be too large to avoid the pile-up in the core and  sufficiently high to maximize signal-to-background ratio in the PSF wings. These conditions are satisfied for instance by a sample of X-ray bright point sources from the \xmm~ Slew Survey \citep[Full Source Catalog, v2.0,][]{2008A&A...480..611S} with the 0.2-2 keV flux from $3\times10^{-12}$ to $10^{-11}$ erg/s/cm$^2$.    

The PSF computed in this way naturally includes averaging over the telescope's field-of-view thanks to the multitude of the source offset angles sampled by the scan patterns. Via stacking, we produce images with 4 arcsec pixel size, which corresponds to the image binning used throughout the paper. In principle, somewhat better characterization of the PSF `core' might be achieved with finer binning of the data, thanks to the sub-pixel algorithm employed in the initial reduction of the raw data \citep{2021A&A...647A...1P}. 

\begin{figure}
\centering
\includegraphics[width=1\columnwidth,bb=40 170 590 690]{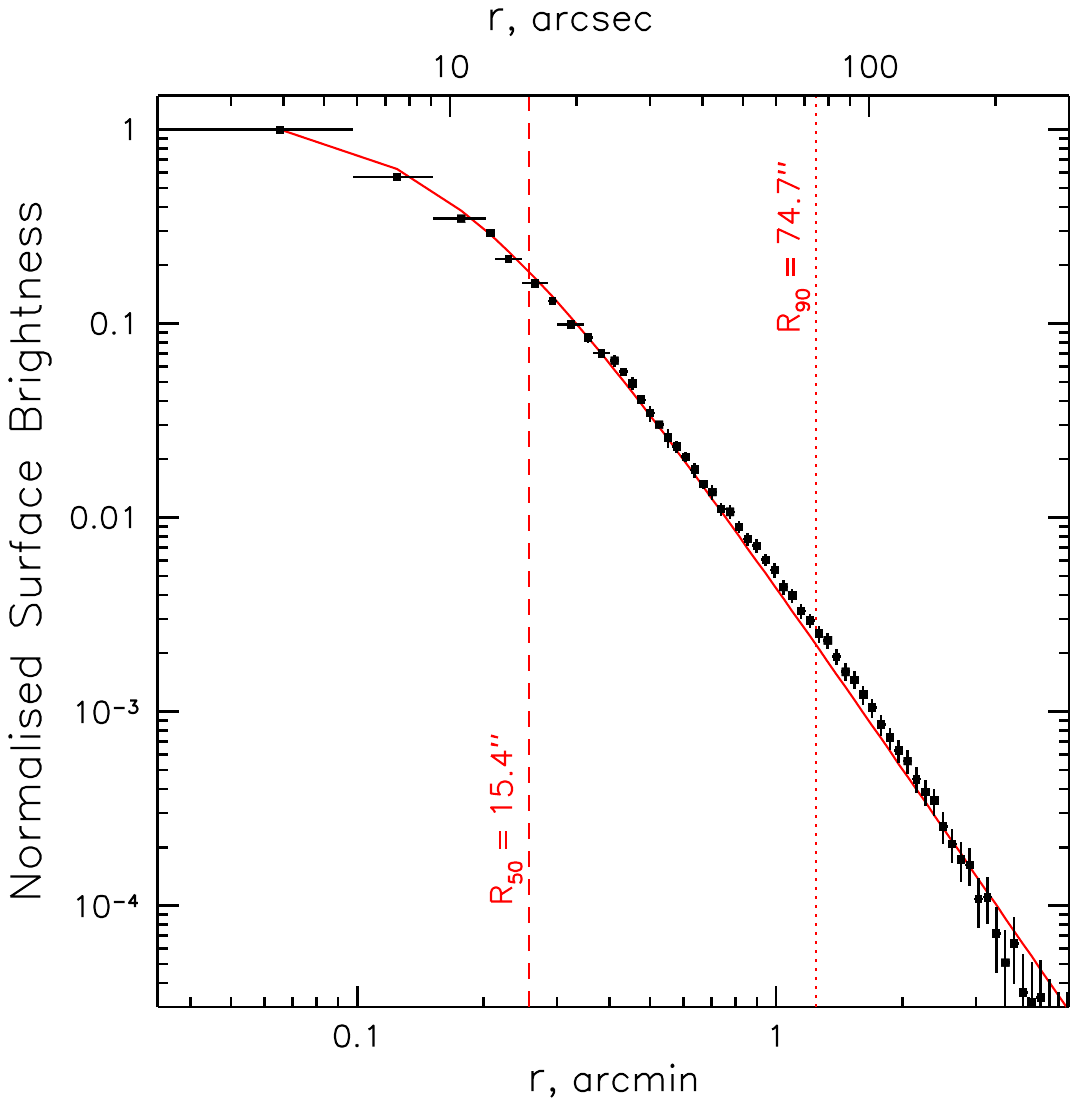}
\caption{The model of the effective Point Spread Function (red curve) used throughout the paper and derived from the stacked images of bright sources  (black points) in 0.4-2 keV energy band. The radii corresponding to 50\% and 90\% of the encompassed photons are marked as $R_{50}$ and $R_{90}$.}
\label{fig:psf_s1}
\end{figure}

For the 0.4-2 keV band, the analysis of the stacked images allowed us to derive a simple approximating model for the effective PSF, namely the beta-model with $\beta\approx0.7$ and $r_{\rm c}\approx10$ arcsec, which is valid up to radial distances of a $\approx4$ arcmin, i.e. within the PSF `core' (see Fig.~\ref{fig:psf_s1}). At large radii, the contribution of the single reflections becomes increasingly noticeable. The derived shape of the `core' PSF slightly differs from the on-ground calibration measurements, although resulting in a very similar Half Power Diameter size, HPD=$2R_{50}\approx30$ arcsec. The aperture encompassing 90\% of the photons has the radius of $R_{90}=\approx75$ arcsec.Throughout the paper, this model is used for the optimal detection and subtraction of the point sources and characterization of the extended emission.

\section{Effective counts production rate for the ICM}
\label{app:epsilon}
\begin{figure}
\centering
\includegraphics[angle=0,width=1.\columnwidth, bb=45 175 585 685]{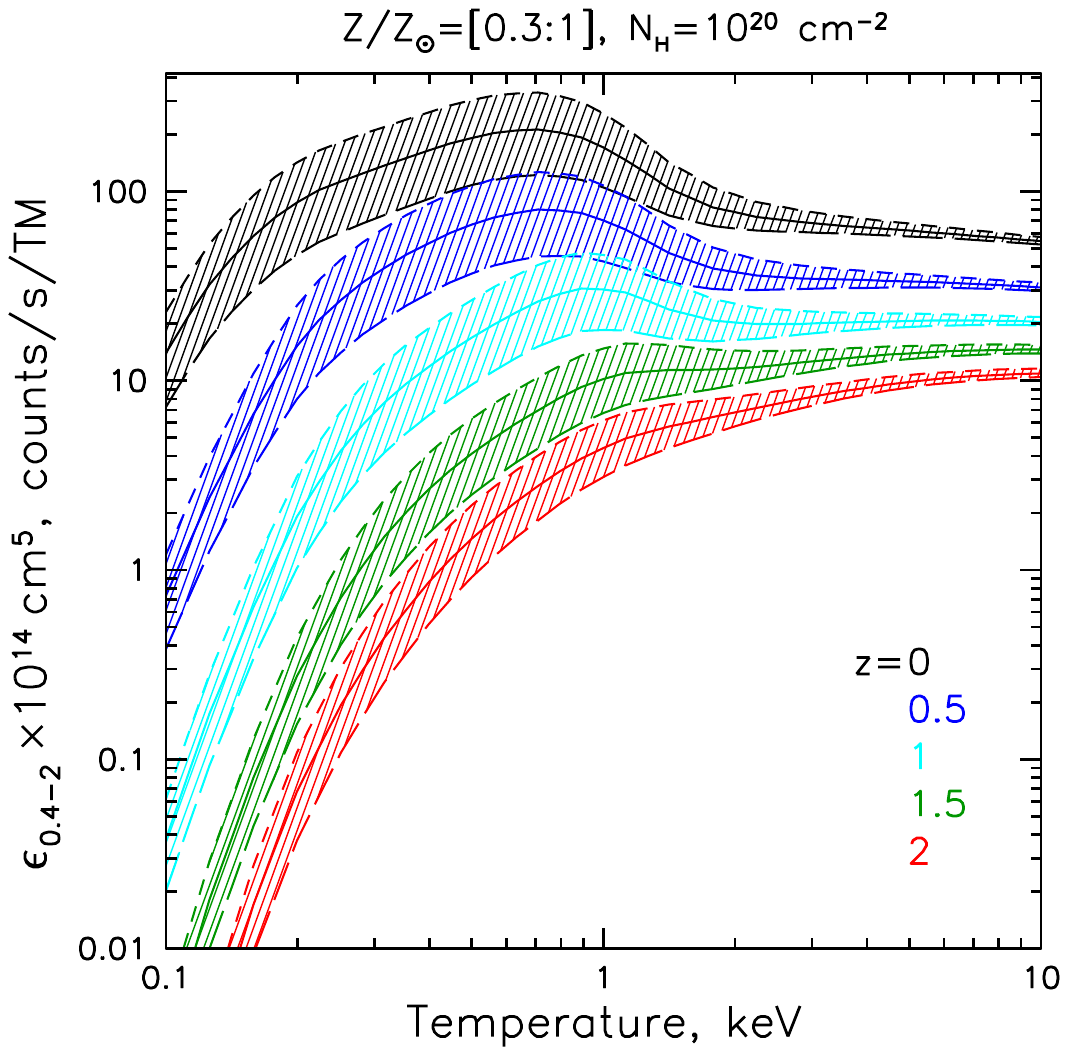}
\caption{Effective ICM emissivity $\epsilon_{c}$ in terms of the 0.4-2 keV count rate per one $SRG$/eROSITA telescope module (with the FOV-averaged effective area) as a function of the ICM temperature. The assumed spectral shape is given by the \texttt{APEC} model at redshifts $z=$ 0 (black), 0.5 (blue), 1(cyan), 1.5 (green), 2 (red) with the metals abundance varying from $Z/Z_{\odot}=0.3$ (long-dashed) to $Z/Z_{\odot}=0.6$ (solid) to $Z/Z_{\odot}=1$ (short-dashed) with respect to the Solar metallicity $Z_{\odot}$. The applied foreground absorption column density equals to $N_{H}=10^{20}$ cm$^{-2}$.}
\label{fig:em_ero}
\end{figure}

The quantity $\epsilon_c$ in its simplest form is a  convolution of the energy resolved thermal plasma emissivity $\Lambda(E,T)$ with the effective area of the instrument $A(E)$. Namely,
\begin{eqnarray}
\epsilon_c=\frac{1}{1+z}\int_{E_1}^{E_2}  \frac{\Lambda(E(1+z),T)}{E } A(E)dE,
\end{eqnarray}
where $E_1$ and $E_2$ are the boundaries of the energy band. In practice, one has to take into account an energy redistribution matrix (as is done in Fig.~\ref{fig:em_ero}). In is well known that for hot ICM, $\epsilon_c$ in the soft band only weakly depends on temperature. This is also true for the 0.4-2 keV band of the eROSITA telescope, as shown in Fig.~\ref{fig:em_ero}. The effective area averaged over the eROSITA FoV was used for this plot. For the temperature range of interest, $\epsilon_c$ varies between $40-55\times 10^{-14} {\rm counts\,s^{-1}\,cm^5}$. 

\section{Comparison of the \textit{ROSAT} and eROSITA radial X-ray surface brightness profiles.}
\label{app:rosat}
\begin{figure}
\centering
\includegraphics[angle=0,trim= 0mm 5cm 0mm 3cm,width=1\columnwidth]{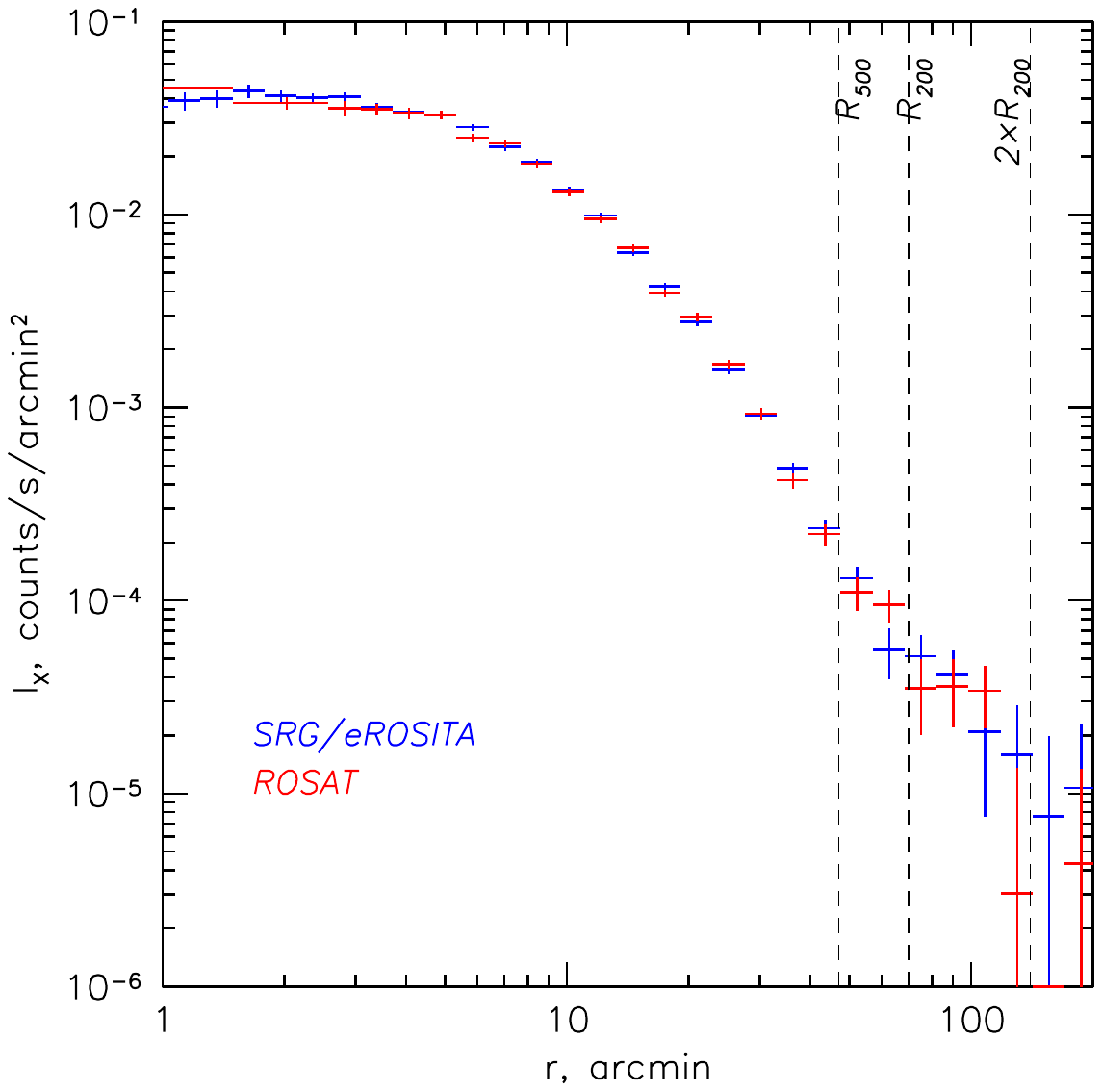}
\caption{Comparison of the \textit{ROSAT} and eROSITA radial profiles of the Coma cluster. In both cases, the data from the all-sky surveys are shown. For \textit{ROSAT} (red points), the data in the 0.4-2.4 keV were used. As in Fig.~\ref{fig:coma_radial}, the 90~degrees wedge to the SW (containing NGC4839 group) has been excluded. The eROSITA points (blue crossed) are the same as in Fig.~\ref{fig:coma_radial}. The \textit{ROSAT} profile has been renormalized to match the eROSITA data. The shapes of the two profiles agree well.}
\label{fig:rass_erass}
\end{figure}

The measured Coma's X-ray surface brightness in the 0.4-2 keV band varies by more than four orders of magnitude over the radial range studied here, and beyond $\sim 30'$ it is smaller than the total background level. To demonstrate that the background subtraction and the account for the contribution of the singly-scattered photons are both corrected for reasonably well, we have compared the shapes of the radial profiles obtained by $SRG$/eROSITA and \textit{ROSAT} in their respective all-sky surveys. For \textit{ROSAT} all-sky survey \citep[e.g.][]{1999A&A...349..389V}, the data in the 0.4-2.4 keV band were used. For comparison with the eROSITA profile, the ROSAT data were renormalized to match at $\sim 10'$. The corresponding profiles are shown in Fig.~\ref{fig:rass_erass}. The agreement is good over the entire range of radii and is fully sufficient for the goals of the present study.

\section{Lorentz factors of electrons providing the largest contribution to the synchrotron emission at a given frequency}
\label{app:gamax}
\begin{figure}
\centering
\includegraphics[angle=0,width=0.95\columnwidth,bb=20 150 600 700]{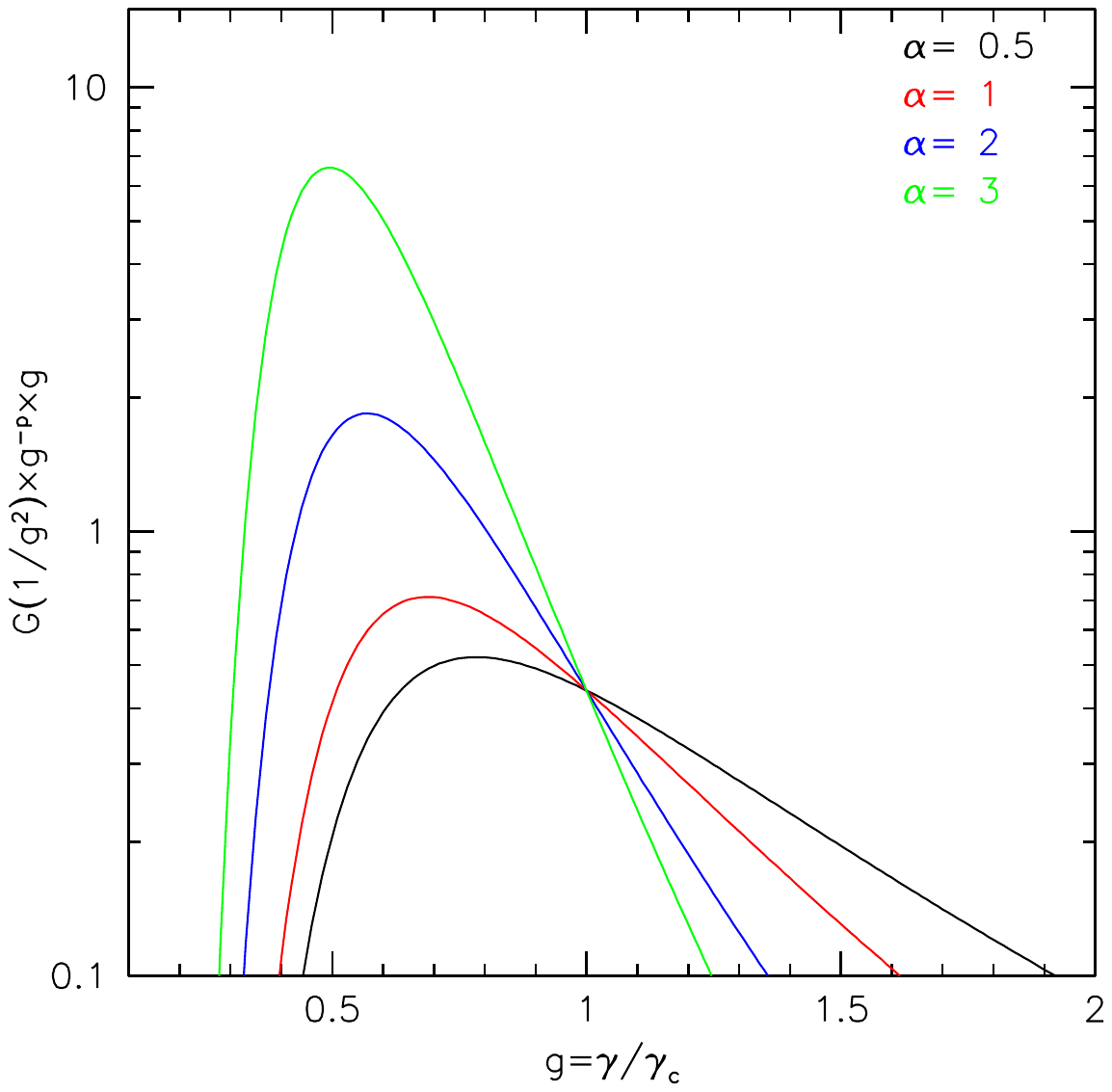}
\caption{Lorentz factor $\gamma$ of electrons providing the most important contribution to the synchrotron emission at a given frequency for a power spectrum of electrons and random orientation of the magnetic field. Here, $\gamma_c$ is set by the condition $\nu=\frac{3}{4\pi}\frac{eB}{m_ec}\gamma^2_c$}
\label{fig:gamax}
\end{figure}

For isotropic electrons with a given Lorentz factor $\gamma$ in a randomly oriented but uniform magnetic field, a useful approximation of the emergent spectrum frequency dependence is given by  \cite{2010PhRvD..82d3002A}, via the function $G(x)$, where $x=\nu/\nu_c$ and $\nu_c=\frac{3}{4\pi}\frac{eB}{m_ec}\gamma^2$. For a power law distribution of electrons $dN/d\gamma\propto\gamma^{-p}$, the maximum contribution to the flux near frequency $\nu$ is provided by electrons with the Lorentz factor $\sim \gamma_{max}$ such, that $F(\gamma)=\gamma^{-p}G(1/\gamma^2)\gamma$ is maximal. It is convenient to use the spectral index $\alpha$ of the observed synchrotron spectrum instead of $p$, where $\alpha=(p-3)/2$ \citep{2011hea..book.....L}. Fig.~\ref{fig:gamax} shows the function $F(\gamma)$ for several values of the spectral index. For our case of interest ($\alpha\sim2$), the function reaches the maximum at $\gamma/\gamma_c\approx 0.57$  \citep[see also][page 164]{1973PhDT.......142C}. This value $\zeta_{max}=\gamma/\gamma_c$ is used in this study to estimate the life time electrons emitting at a given frequency.

\label{lastpage}

\end{document}